\documentclass[a4paper,fleqn,usenatbib]{mnras}

\usepackage[T1]{fontenc}
\usepackage{ae,aecompl}

\usepackage{graphicx}	
\usepackage{amsmath}	
\usepackage{amssymb}	
\usepackage{bm}

\usepackage[normalem]{ulem}
\usepackage[usenames,dvipsnames,svgnames,table]{xcolor}

\newcommand{\rkl}[1]{\left(#1\right)}
\newcommand{\ekl}[1]{\left[#1\right]}

\newcommand{\skl}[1]{\left\langle#1\right\rangle}

\newcommand{\slope}{q}

\newcommand{\vektor}[1]{\bm{\mathit{#1}}}
\newcommand{\tensor}[1]{\bm{\mathsf{#1}}}
\newcommand{\vnabla}{\bm{\nabla}}
\newcommand{\bcdot}{\bm{\cdot}}
\newcommand{\dd}{\rmn{d}}

\title[Spectral CR-MHD]{Spectrally resolved cosmic ray hydrodynamics -- I. Spectral scheme}

\author[Philipp Girichidis et al.]{Philipp~Girichidis$^{1}$\thanks{E-mail: philipp@girichidis.com}, Christoph Pfrommer$^{1}$, Micha{\l}~Hanasz$^{2}$, Thorsten Naab$^{3}$\\
$^1$Leibniz-Institut f\"{u}r Astrophysik Potsdam (AIP), An der Sternwarte 16, 14482 Potsdam, Germany\\
$^2$Centre for Astronomy, Nicolaus Copernicus University, Faculty of Physics, Astronomy and Informatics, Grudziadzka 5, PL-87100 Toru\'n, Poland\\
$^3$Max-Planck-Institut f\"{u}r Astrophysik, Karl-Schwarzschild-Str. 1, 85741 Garching, Germany}

\date{Accepted XXX. Received YYY; in original form ZZZ}

\pubyear{2018}

\begin{document}
\begin{NoHyper}
\label{firstpage}
\pagerange{\pageref{firstpage}--\pageref{lastpage}}
\maketitle

\begin{abstract}
Cosmic ray (CR) protons are an important component in many astrophysical systems. Processes like CR injection, cooling, adiabatic changes as well as active CR transport through the medium strongly modify the CR momentum distribution and have to be taken into account in hydrodynamical simulations. We present an efficient novel numerical scheme to accurately compute the evolution of the particle distribution function by solving the Fokker-Planck equation with a low number of spectral bins ($10 - 20$), which is required to include a full spectrum for every computational fluid element. The distribution function is represented by piecewise power laws and is not forced to be continuous, which enables an optimal representation of the spectrum. The Fokker-Planck equation is solved with a two-moment approach evolving the CR number and energy density. The low numerical diffusion of the scheme reduces the numerical errors by orders of magnitude in comparison to classical schemes with piecewise constant spectral representations. With this method not only the spectral evolution of CRs can be computed accurately in magnetohydrodynamic simulations but also their dynamical impact as well as CR ionisation. This allows for more accurate models for astrophysical plasmas, like the interstellar medium, and direct comparisons with observations.
\end{abstract}





\section{Introduction}

CRs are an important energy component in many astrophysical systems from proto-planetary discs, to the interstellar medium in galaxies, and to galaxy clusters \citep{StrongMoskalenkoPtuskin2007,GrenierBlackStrong2015}. Of particular importance are CR protons in galaxies because their energy densities are comparable to magnetic, thermal and kinetic energy densities and because of their resulting dynamical and chemical impact on the gas. Early theoretical models  \citep{Krymskii1977, AxfordLeerSkadron1977, Bell1978, BlandfordOstriker1978} highlight the acceleration of CRs at strong shocks via diffusive shock acceleration, which has been successfully modelled numerically \citep{CaprioliSpitkovsky2014a}. For Galactic CRs the most abundant shocks are supernova remnants with evidence of hadronic particle acceleration \citep{ZirakashviliAharonian2010, MorlinoCaprioli2012, AckermannEtAl2013}, see also \citet{Blasi2013} and \citet{Amato2014}.

The coupling of CRs to the gas is mediated via the generation of Alfv\'{e}n waves and the resulting scattering. The simplest way to describe CRs as a fluid is the one-moment approach in the scattering angle $\mu$ under the assumption that the particle distribution function is isotropic \citep[see, e.g.][]{Zweibel2013}. A further simplification for numerical models is a grey approach, in which the distribution function times the kinetic energy per particle is integrated over momentum space and the resulting CR energy density is evolved in time and space. Recently, \citet{JiangOh2018} and \citet{ThomasPfrommer2019} extended the traditional one-moment schemes in the scattering angle to two-moment descriptions that include a more self-consistent coupling of CRs to the plasma and captures streaming and diffusion relative to the gas rest frame.

Previous studies have modelled the dynamical impact of CR protons in the galactic ISM by including them as a relativistic fluid with an effective adiabatic index, see e.g. \citet{NaabOstriker2017}. The first dynamical coupling was performed by \citet{HanaszLesch2003} studying the Parker instability. A prominent application of CR hydrodynamics is their role in driving galactic outflows \citep{Ipavich1975, Breitschwerdt1991, Breitschwerdt1993, PtuskinEtAl1997, Everett2008, SocratesDavisRamirezRuiz2008, SamuiSubramanianSrianand2010, DorfiBreitschwerdt2012, RecchiaBlasiMorlino2016}, but see also applications in galaxy clusters \citep{BlasiColafrancesco1999, RuszkowskiYangReynolds2017,EhlertEtAl2018}. Previous hydrodynamical simulations used global disc setups \citep{JubelgasEtAl2008, UhligEtAl2012, HanaszEtAl2013, BoothEtAl2013, SalemBryan2014, PakmorEtAl2016, PfrommerEtAl2017, JacobEtAl2018} with a focus on the large-scale dynamics as well as stratified boxes of a representative fractions of the ISM \citep{GirichidisEtAl2016a, SimpsonEtAl2016, FarberEtAl2018, GirichidisEtAl2018a} with a focus on the chemical evolution, the detailed CR coupling and the relative importance between CRs and SNe as a driver.

All of the previous studies reveal that CRs can provide relevant pressures and accelerations of the gas. However, to what extent they provide a main contribution to the dynamical evolution depends on the system under consideration and the details of the CR parameters. In particular the combination of CR losses and their spatial transport might be important in determining the global impact. Both processes are strong functions of the spectral energy distribution of CRs, which is not resolved in current CR-MHD simulations but only integrated to yield a total CR energy \citep[with the notable exception of the simplified spectral treatment of][]{PfrommerEtAl2006,EnsslinEtAl2007,JubelgasEtAl2008}.

The dynamical work of CRs on the gas -- and vice versa -- results in adiabatic changes that are connected with compression and expansion of the gas. In addition low-energy CRs lose energy via Coulomb collisions with the thermal particles of the gas. Collisions of high-energy CRs at energies above GeV with the gas result in catastrophic hadronic losses via the production of neutral pions and their decay into $\gamma$-ray photons. Strong shocks further accelerate CRs. All together the CR energy and the spectral distribution is constantly changing.

In the interstellar medium the efficient coupling of CRs with the gas via magnetic fields results in a non-negligible effective CR pressure that thickens the galactic disc and launches outflows from the galaxies \citep[e.g.][]{GirichidisEtAl2016a,SimpsonEtAl2016,FarberEtAl2018}. This direct dynamical impact is mainly due to CRs with momenta of a few GeV/$c$. Low-energy CRs suffer from strong Coulomb losses, which reduces their energy density and results in a negligible impact via their pressure. However, a perceptible increase of the cross section of MeV-to-GeV CRs with the thermal gas causes efficient ionisation of the gas \citep{Dalgarno2006, Padovani2009, IvlevEtAl2018, PhanMorlinoGabici2018}. As CRs can penetrate deeply into dense molecular clouds, they influence the formation of stars and the observational signatures.

Above a total energy of $E_\mathrm{thr}=1.22\,\mathrm{GeV}$ CRs are energetic enough to produce pions, which in turn decay into $\gamma$-ray photons, secondary electrons and neutrinos. The secondary electrons can emit radio synchrotron emission in ubiquitous magnetic fields and Compton upscatter ambient radiation fields into the X-ray to gamma-ray regime. A predictive modelling of the resulting non-thermal emission processes calls for self-consistent spectral modelling of the CR spectrum in time and space. While the CR spectrum, its composition, and the non-thermal radiative signatures are modelled in CR propagation codes \citep{StrongMoskalenko1998, Kissmann2014, EvoliEtAl2017}, such approaches assume the Galaxy to be static and adopt observationally inferred distributions of the gas density, magnetic fields, and CR sources that are not necessarily emerging from a self-consistent simulation of a dynamically evolving galaxy.

As CRs have a dynamical impact it is thus favourable to follow the spectrum together with the hydrodynamical evolution, i.e. to compute a full spectrum for every computational cell. This poses strong constraints on the numerical scheme. We would like to follow a large dynamical range in CR energy from below MeV to above TeV. The CR spectrum itself is very steep, i.e. covers a large dynamical range in amplitude. Nonetheless we can only represent the spectrum with a low number of spectral bins ($\sim10-30$), which requires a relatively complex numerical scheme compared to standard methods with orders of magnitude larger spectral resolution.

In a series of papers we introduce a novel implementation of the spectral CR energy distribution, which allows a dynamical coupling of CRs with the gas as well as an accurate evolution of the CR spectrum for the relevant mechanism by only using a low number of spectral bins compared to classical spectral approaches. In Section~\ref{sec:background} we present the theoretical background and present analytical solution to idealised cases. In Section~\ref{sec:spectral-discretisation} we outline the spectral discretisation of the particle distribution function. In Section~\ref{sec:derivation} we derive a numerical scheme for the time evolution of the CR spectrum and show one-zone tests of it in Section~\ref{sec:one-zone}. We present a one-dimensional test of energy dependent spatial diffusion in Section~\ref{sec:diffusion} and conclude in Section~\ref{sec:conclusions}.

\section{Evolution of CRs}
\label{sec:background}

Before presenting the discretisation scheme and the numerical algorithms, we review the time evolution of CRs including the individual loss processes as well as the combined solutions of freely cooling and the steady state spectrum for continuous injection, which we will use to compare and scrutinise our numerical simulations.

\subsection{Theoretical background}

Cosmic rays are charged particles and therefore interact with the magnetic field. Quasi-linear theory in the frequent scattering limit leads to the Fokker-Planck equation for the phase space particle distribution function \citep{Schlickeiser1989,Miniati2001}. Throughout the paper we use the three-dimensional form of the distribution function, $f=f^{(3)}$. Other studies \citep{EnsslinEtAl2007,WinnerEtAl2019} use the equivalent one-dimensional form $f^{(1)}$, where $f^{(1)} = 4\pi p^2 f$. The Fokker-Planck equation then reads
\begin{align}
  \frac{\partial f}{\partial t} = & \underbrace{-\vektor{u}\bcdot\vnabla f}_{\text{advection}} + \underbrace{\vnabla\bcdot\rkl{\tensor{D}_{xx}\bcdot \vnabla f}}_{\text{diffusion}} + \underbrace{\frac 1 3 \rkl{\vnabla\bcdot\vektor{u}}p\frac{\partial f}{\partial p}}_{\text{adiabatic process}}\nonumber\\
  & + \underbrace{\frac{1}{p^2}\frac{\partial}{\partial p}\ekl{p^2\rkl{b_l f + D_{pp}\frac{\partial f}{\partial p}}}}_{\text{other losses and Fermi II acceleration}} + \underbrace{j}_{\text{sources}}\label{eq:FP},
\end{align}
where $f = f(\vektor{x},\vektor{p},t)$ is the isotropic part of the distribution function and $\tensor{D}_{xx}$ and $D_{pp}$ are the spatial diffusion tensor and the momentum space diffusion coefficient, respectively. The losses are described as $b_l(\vektor{x},\vektor{p},t)=\dd p/\dd t$. We note that formally only continuous losses can be cast into this form but not impulsive losses like, e.g. hadronic losses because they do not conserve the number of particles \citep{Schlickeiser2002}. However, we will be taking moments of equation~\eqref{eq:FP}, which implies integrating over the distribution function  that consists of a large population of individual particles. It is therefore justified to take the continuous limit of the collection of
these interactions. The term $j(\vektor{x},\vektor{p},t)$ describes CR sources. Here, we neglect CR streaming for simplicity \citep[see][for including this process in a grey approach]{JiangOh2018,ThomasPfrommer2019}. Because second-order Fermi acceleration is a slow process that only mildly shifts the CR spectrum toward larger momenta, we postpone a treatment of turbulent reacceleration to future work and set $D_{pp}=0$. For the remainder of the paper we therefore only discuss spatial diffusion ($\tensor{D}_{xx}$) and omit the subscript $xx$. Generally, the distribution function and all its components depend on the position in space, $\vektor{x}$ and time, $t$. In order to simplify the equations, we omit these dependencies unless explicitly needed.

The number density $n$ and energy density $e$ are then given by the appropriate moments of $f$,
\begin{align}
  n_\mathrm{CR} &= \int_0^\infty 4\pi p^2 f(p) \,\dd p\\
  e_\mathrm{CR} &= \int_0^\infty 4\pi p^2 f(p) T(p)\,\dd p.
\end{align}
Here, $T(p)$ is the kinetic energy of the CRs,
\begin{align}
T(p) =& \sqrt{p^2c^2 + m_\mathrm{p}^2c^4} - m_\mathrm{p} c^2,
\end{align}
where $m_\mathrm{p}$ is the proton mass and $c$ is the speed of light. The CR pressure is given by
\begin{align}
  P_\mathrm{CR} &= \int_{0}^{\infty} \frac{4\pi}{3}\, c\,p^3 \beta(p) f(p) \dd p\notag\\
  &= \int_{0}^{\infty} \frac{4\pi}{3}\, \frac{ f(p)\,p^4c^2}{\sqrt{m^2c^4+p^2c^2}}\dd p
\end{align}
where $\beta(p) = p/\sqrt{p^2+(m_\mathrm{p}c)^2}$.

\subsection{Adiabatic changes}

We start with the simplest process, which is the adiabatic one,
\begin{equation}
\label{eq:ad-process}
\rkl{\frac{\partial f}{\partial t}}_{\mathrm{ad}} = -\rkl{\frac{1}{3}\vnabla\bcdot\vektor{u}}\,\frac{\partial f}{\partial \ln p}.
\end{equation}
The divergence of the velocity field $\vnabla\bcdot\vektor{u}$ is constant over a spectral time integration because we apply operator splitting for the hydrodynamics and the spectral evolution. We note that the adiabatic process is simply equivalent to an advection in logarithmic space with the advection speed $-1/3\,\vnabla\bcdot\vektor{u}$. This means that the local slope in a comoving frame $\vnabla\bcdot\vektor{u}$ does not change. Using
\begin{equation}
\frac{\partial f}{\partial t} = \frac{\partial f}{\partial p}\,\frac{\dd p}{\dd t}
\end{equation}
in equation~(\ref{eq:ad-process}) we find 
\begin{equation}
\label{eq:adiabatic-compression-0}
\frac{\dd p}{\dd t} = - \rkl{\frac{1}{3}\vnabla\bcdot\vektor{u}} p.
\end{equation}
Separation of variables and integrating from $p (t_0)$ to $p(t)$, which corresponds to the temporal evolution from $t_0$ to $t=t_0+\Delta t$, yields
\begin{equation}
\label{eq:adiabatic-compression}
p(t) = p(t_0) \,\exp\ekl{-\int_{t_0}^{t_0+\Delta t}\rkl{\frac{1}{3}\vnabla\bcdot\vektor{u}}\,\dd t}.
\end{equation}

\subsection{Coulomb losses}

\begin{figure}
  \centering
  \includegraphics[width=8cm]{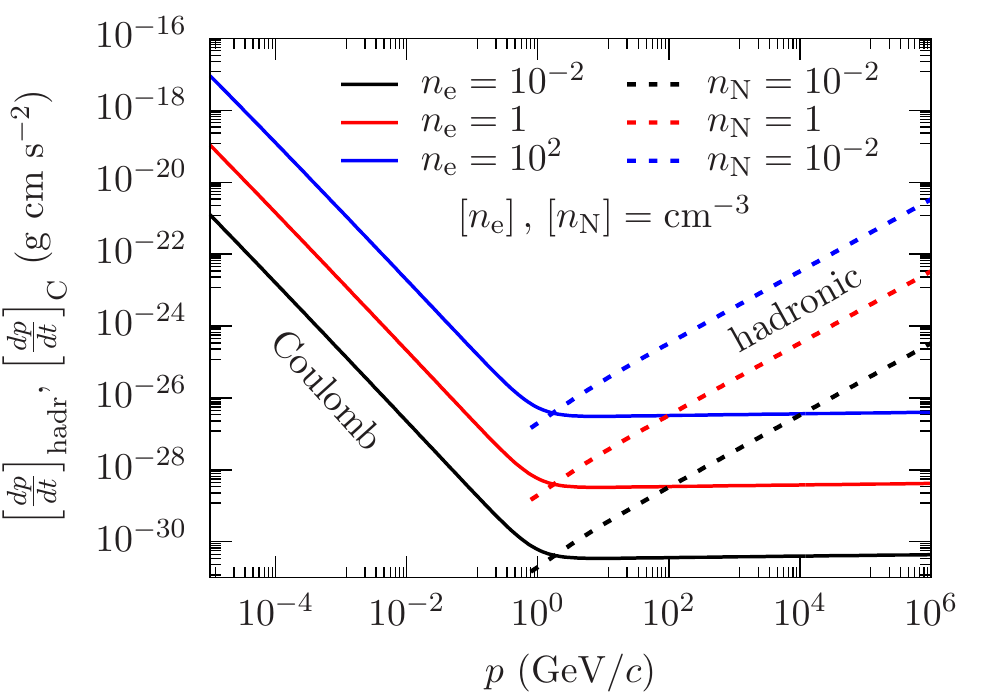}
  \caption{Coulomb and hadronic losses as a function of momentum for different target densities, where $n_\mathrm{e}$ and $n_\mathrm{N}$ are the electron and nucleon densities. Coulomb losses dominate at low CR momenta with a scaling close to $p^{-1.9}$. Hadronic losses start at the threshold momentum or pion production of $p_\mathrm{thr}\approx0.78\,\mathrm{GeV}/c$ and scale linearly with $p$.}
  \label{fig:coulomb-hadronic-losses}
\end{figure}

\begin{figure}
  \centering
  \includegraphics[width=8cm]{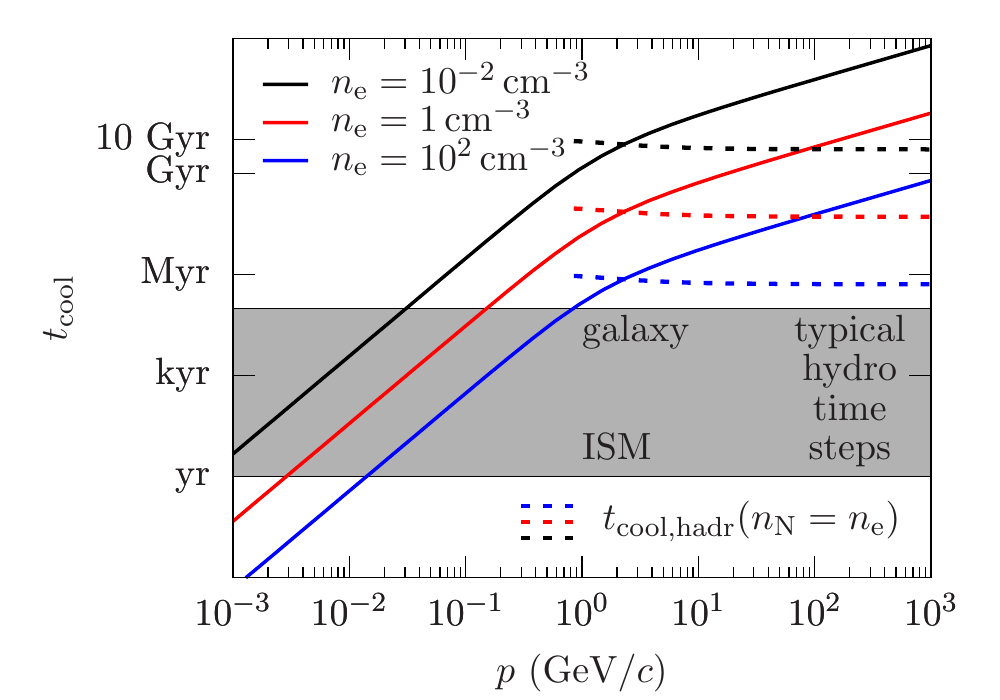}
  \caption{Cooling time as a function of the initial momentum for different electron densities. For low CR momenta the Coulomb cooling times are significantly shorter than typical hydrodynamical times. At momenta above $\sim1\,\mathrm{GeV}/c$ the hadronic time-scales are shorter than the Coulomb cooling times but larger than typical hydrodynamical time steps.}
  \label{fig:cooling-times}
\end{figure}

The total energy loss per proton is \citep{Gould1972}
\begin{align}
\label{eq:coulomb-losses}
  -\left[\frac {\dd T(p)}{\dd t}\right]_\rmn{C} &= \frac{\omega_\mathrm{pl}^2 e^2}{\beta c}\left[ \ln\left( \frac{2m_\mathrm{e} c^2\beta p}{\hbar \omega_\mathrm{pl} m_\mathrm{p}c}\right) - \frac{\beta^2}{2}\right],
\end{align}
where $\omega_\mathrm{pl}=\sqrt{4\pi e^2 n_\mathrm{e}/m_\mathrm{e}\,}$ is the plasma frequency and $n_\mathrm{e}$ is the electron number density. Coulomb and hadronic losses are shown as a function of momentum for different electron densities in Fig.~\ref{fig:coulomb-hadronic-losses}. We note that the weak scaling with $n_e$ in the logarithmic term can be neglected in comparison to the linear scaling of the Coulomb loss term ($\propto\omega_\mathrm{pl}$). To illustrate the scaling we can express the losses as a function of $p$ instead of $T$ and can use the simplified approximation
\begin{align}
	b_\mathrm{C} &\equiv \left[\frac{\dd p}{\dd t}\right]_\mathrm{C}\\
	\label{eq:coulomb-loss-approximate}
	&\approx  -10^{-18}\,\mathrm{erg\,cm^3\,s^{-1}}\frac{n_\mathrm{e}}{c}\left[1+\left(\frac{p}{\mathrm{GeV}/c}\right)^{-1.9}\right],
\end{align}
which is accurate to $17\%$ over the range shown. The scaling with $p^{-1.9}$ for low momenta leads to a finite cooling time, at which the momentum reaches zero. The cooling times to $p=0$ are plotted as a function of initial momentum in Fig.~\ref{fig:cooling-times} for different electron densities. Typical hydrodynamical time steps ($\Delta t_\mathrm{hydro}$) are indicated by the grey area, which illustrates that for momenta below $1\,\mathrm{GeV}/c$ the hydrodynamical time step might be larger than the cooling time. For $t_\mathrm{cool}\lesssim \Delta t_\mathrm{hydro}$ we can use the numerical cooling scheme. If cooling occurs on much shorter time-scales compared to the simulation time step we can directly evaluate the steady state solution without intensive numerical integration. For the tests in this paper we can set the time steps independently of any hydrodynamical simulation and therefore use the fully numerical solution.

\subsection{Hadronic losses}
\label{sec:hadronic-losses}

Hadronic losses occur as a result of inelastic reactions of CRs with the gas. The interactions mainly result in the production of pions if the CR energy exceeds the threshold momentum, $p_\mathrm{thr}\approx0.78\,\mathrm{GeV}/c$. The loss rate of kinetic energy is given by
\begin{align}
-\rkl{\frac{\dd T}{\dd t}} &= c\,n_\mathrm{N}\sigma_\mathrm{pp}K_\mathrm{p}T(p)\theta(p-p_\mathrm{thr}).
\end{align}
Here, $n_\mathrm{N}$ is the target nucleon density of the ISM, $\sigma_\mathrm{pp}$ is the total pion cross section and $K_\mathrm{p}\approx 1/2$ is the inelasticity of the reaction \citep{MannheimSchlickeiser1994}. The CR momentum losses are shown in Fig.~\ref{fig:coulomb-hadronic-losses}. The cooling times for hadronic losses are shown in Fig.~\ref{fig:cooling-times}. We note that the typical cooling times for hadronic losses are long compared to typical hydrodynamical time steps. We note that the losses scale with the kinetic energy, $T(p)$, which asymptotes to a linear scaling in momentum, $p$, for relativistic CRs. As a consequence, the spectral changes due to hadronic losses in the relativistic case are
\begin{align}
\rkl{\frac{\partial f}{\partial t}}_\mathrm{hadr} &\propto \frac{1}{p^2}\,\frac{\partial}{\partial p} \big(p^2 b_\mathrm{hadr}f\big),
\end{align}
where
\begin{equation}
b_\mathrm{hadr} = \frac{\dd p}{\dd T}\,\left(\frac{\dd T}{\dd t}\right)_\mathrm{hadr}.
\end{equation}
If $f$ is a power-law in $p$, we note that the hadronic losses have the same scaling with momentum as $f$,
\begin{align}
\rkl{\frac{\partial f}{\partial t}}_\mathrm{hadr} &\propto f,
\end{align}
which leaves the slope of the particle distribution function unchanged.

\subsection{Fermi-I acceleration}

Fermi-I or diffusive shock acceleration is an important gain process of energy, in which thermal particles can be accelerated to super-thermal energies by means of adiabatic compression and expansion together with spatial diffusion at shocks \citep{Bell1978, Drury1983, BlandfordEichler1987}. In principle both processes are included in the Fokker-Planck equation, so that Fermi-I acceleration can be accounted for analytically. However, a numerical solution requires resolutions of the kinetic physics, in particular the scattering of particles due to the non-resonant hybrid instability on small spatial scales around the shock front \citep{Bell2004}, which is impossible to resolve in our applications of astrophysical fluid dynamics. We therefore need to treat the acceleration of CRs as a subgrid model and effectively describe it as an injection of CR energy in regions of strong shocks. Given the shock compression ratio $r=\rho_\mathrm{post}/\rho_\mathrm{pre}$, with the pre and post-shock densities $\rho_\mathrm{pre}$ and $\rho_\mathrm{post}$, we expect injection with a spectral index
\begin{equation}
\slope_\mathrm{acc} = \frac{3r}{r-1}.
\end{equation}
The effective Fermi-I acceleration is thus encoded in the sources $j$ with
\begin{equation}
j_\mathrm{acc} = A_\mathrm{acc} p^{-\slope_\mathrm{acc}}\exp(-p/p_\mathrm{acc}),
\end{equation}
where the acceleration efficiency and thus $A_\mathrm{acc}$ and the maximum momentum $p_\mathrm{acc}$ depends on the local shock conditions \citep{Bell2013}, see also \citet{MarcowithEtAl2016}.

\subsection{Free cooling}
\label{sec:free-cooling}

\subsubsection{General theoretical considerations for the spectral behaviour}

We can investigate the expected spectral changes by assuming conservation of the total CR number density in the same volume,
\begin{align}
\label{eq:particle-number-conservation}
n(t_1) &= n(t_0),\\
\label{eq:particle-number-conservation-diff}
4\pi\int_{p_1}^{\infty}p^2\,f(p,t_1)\dd p &= 4\pi\int_{p_0}^{\infty}p'^2\,f(p',t_0)\dd p'.
\end{align}
Here, $p_0$ and $p_1$ are related via the change in momentum over time, which we address below. Assuming that $f(\infty)$ vanishes, i.e. the number density of CRs is finite, we can assume conservation of the number density in an infinitesimal momentum interval to find
\begin{align}
f(p_1,t_1) &= f(p_0,t_0)\,\frac{p_0^2}{p_1^2}\,\frac{\dd p_0}{\dd p_1}\label{eq:distr-function-time-evol},
\end{align}
where the differential changes in $p$, $\dd p$, depend on the momentum and thus on the time, such that we can write
\begin{align}
\frac{\dd p_0}{\dd p_1} &= \frac{\dd p}{\dd t}(p_0) \ekl{\frac{\dd p}{\dd t}(p_1)}^{-1}.
\end{align}
The changes in momentum are described by the generalised loss term
\begin{align}
\ekl{\frac{\dd p}{\dd t}}_\mathrm{X} &= b_\mathrm{X}.
\end{align}
where we assume a simple power-law scaling with momentum $p$
\begin{align}
\ekl{\frac{\dd p}{\dd t}}_\mathrm{X} &= b_\mathrm{X,0} p^h,\quad h\neq-1 \quad\text{for simplicity}.
\end{align}
We solve the differential equation by separation of variables and integration
\begin{align}
\int_{p_0}^{p_1}\,\frac{\dd p'}{b_\mathrm{X}(p)} &= (t_1-t_0)
\end{align}
and we can solve for the momentum $p_1$
\begin{align}
p_1 = \ekl{p_0^{1-h} + (1-h)b_\mathrm{X,0}(t_1-t_0)}^{1/(1-h)}. 
\end{align}
For losses $b_\mathrm{X,0} < 0$, $p_1<p_0$ and depending on $h$ the momentum will cool to $p_1=0$ within a finite time. Inserting the momentum as well as the loss rates into equation~(\ref{eq:distr-function-time-evol}) yields
\begin{align}
f(p_1,t_1) &= f(p_0,t_0)\,\frac{p_0^2}{p_1^2}\,\frac{p_0^h}{p_1^h}.
\end{align}

\subsubsection{Application to approximate Coulomb cooling}

Coulomb cooling for low momenta scales as $p^{-1.9}$, i.e. $h=-1.9$. Hence, we find
\begin{align}
f(p_1,t_1) &= f(p_0,t_0)\,p_0^{0.1}p_1^{-0.1}\notag\\
&=f(p_0,t_0)\,p_0^{0.1}\ekl{p_0^{2.9} + 2.9\,b_\mathrm{C,0}(t_1-t_0)}^{-0.1/2.9}.
\end{align}
As long as we do not cool the momentum to zero we can assume that $p_0^{2.9} > 2.9\,b_\mathrm{C,0}(t_1-t_0)$. For $p_0^{2.9} \gg 2.9\,b_\mathrm{C,0}(t_1-t_0)$ we can neglect the second term in the sum and find again
\begin{align}
f(p_1,t_1) &= f(p_0,t_0),
\end{align}
a vanishing scaling with $p$, i.e. a flat slope.

\subsection{Steady state spectrum}
\label{sec:steady-state}

Here we focus on the steady state solution resulting from cooling and continuous CR injection. The left-hand side of Equation~(\ref{eq:FP}) thus vanishes and the resulting equation reads
\begin{align}
\frac{\partial f}{\partial t} & = 0 = \frac{1}{p^2}\frac{\partial}{\partial p}\ekl{p^2 b_l(p) f(p)} + j(p),
\end{align}
which can be rewritten to yield
\begin{align}
f(p) &= -\frac{1}{p^2 b_l(p)}\int_{p}^{\infty}p^2 j(p) \dd p.
\end{align}
The losses include hadronic and Coulomb interactions
\begin{align}
b_l(p) & = \rkl{\frac{\dd p}{\dd t}}_\mathrm{tot} = \ekl{\rkl{\frac{\dd p}{\dd t}}_\mathrm{hadr}+\rkl{\frac{\dd p}{\dd t}}_\mathrm{Coul}}.
\end{align}
The injection $j(p)$ is modelled as a power-law spectrum with
\begin{align}
j(p) &= Ap^{-\slope}.
\end{align}
The steady state solution converges for $\slope>3$ and the steady state solution reads
\begin{align}
  f(p) &= \frac{A\,p^{-\slope+1}}{(\slope-3)\,b_l(p)}
  \label{eq:steady-state}
\end{align}

\section{Spectral discretisation of the particle distribution function}
\label{sec:spectral-discretisation}

\begin{figure}
\includegraphics[width=8cm]{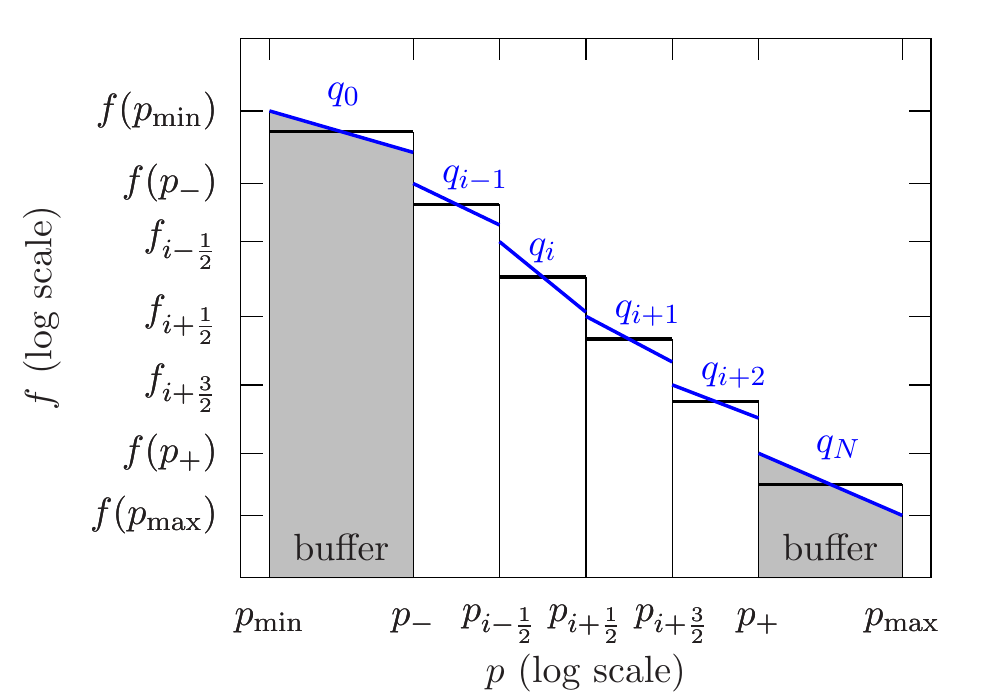}
\caption{Illustration of the spectral discretisation. We use a piecewise power-law representation of the CR spectrum with local amplitudes $f_{i-1/2}$ and slopes $\slope_i$ (blue lines), which is more accurate than a piecewise constant representation, in particular for a low number of momentum bins and steep spectra. The spectrum is not forced to be continuous. At the low and high momentum end of the spectrum we use a larger buffer bin.}
\label{fig:spectral-grid-sketch}
\end{figure}

\subsection{Discretization in momentum}

In principle we can evolve the Fokker-Planck equation using $f$ directly, similar to previous approaches as in \textsc{GALPROP} \citep{StrongMoskalenko1998}, \textsc{PICARD} \citep{Kissmann2014}, \textsc{DRAGON2} \citep{EvoliEtAl2017} or \textsc{CREST} \citep{WinnerEtAl2019}. However, a standard discretisation with piecewise constant values for $f$ requires relatively high spectral resolution in order to obtain accurate results \citep[see discussion in][]{WinnerEtAl2019}. Reasonably low errors are achieved with approximately 50-100 bins per momentum decade, i.e. much more than a few hundred bins for a full spectrum ranging from $p\sim10^{-1}$ to $100\,\mathrm{GeV}/c$. This high number stems from the fact that we need to cover a large range in momentum and  -- due to the steep CR spectra -- a large dynamical range of $f$.

For our CR module we focus on CR protons and their dynamical impact on the hydrodynamical evolution. This requires to solve the spectrum in every hydrodynamical cell and evolving hundreds of momentum bins is not feasible. Instead we aim for a relatively low number of spectral bins of order $10-20$. As a consequence, we need a numerical scheme, which can accurately treat the large dynamic range with only several cells and does not suffer from strong numerical diffusion. We therefore chose a logarithmic spacing for the spectral discretisation and describe the particle distribution function as piecewise power laws.

The spectral distribution in principle covers the entire momentum space. We restrict our computation to a finite range $p_\mathrm{min}<p<p_\mathrm{max}$. We discretise the spectral distribution with $N_\mathrm{bins}$ bins between $p_\mathrm{min}$ and $p_\mathrm{max}$ as illustrated in Fig.~\ref{fig:spectral-grid-sketch}. The logarithmically spaced bins include two buffer or boundary bins at the lower and the upper end of the spectrum, where the lower buffer bin ranges from $p_\mathrm{min}$ to $p_-$ and the upper one from $p_+$ to $p_\mathrm{max}$. The bins between $p_-$ and $p_+$ are equally spaced in $\log p$. We define cell centred quantities with index $i$ and corresponding left-hand cell-faced quantities with index $i-1/2$. We adopt a distribution function $f$ as a piecewise power-law
\begin{equation}
f(p) = f_{i-1/2} \rkl{\frac{p}{p_{i-1/2}}}^{-\slope_i},
\end{equation}
with the slopes $\slope_i$ (blue lines in Fig.~\ref{fig:spectral-grid-sketch}). We note that this functional approximation of $f$ has two degrees of freedom per bin, namely the amplitude $f_{i-1/2}$ and the slope $\slope_i$. We therefore also investigate two moments of $f$ per bin, i.e. number and energy density, which are given by
\begin{align}
  n_i &= \int_{p_{i-1/2}}^{p_{i+1/2}} 4\pi p^2 f(p) \,\dd p\\
  e_i &= \int_{p_{i-1/2}}^{p_{i+1/2}} 4\pi p^2 f(p) T(p)\,\dd p.
\end{align}

\subsection{Motivation for a two-moment approach}

\begin{figure}
\includegraphics[width=8cm]{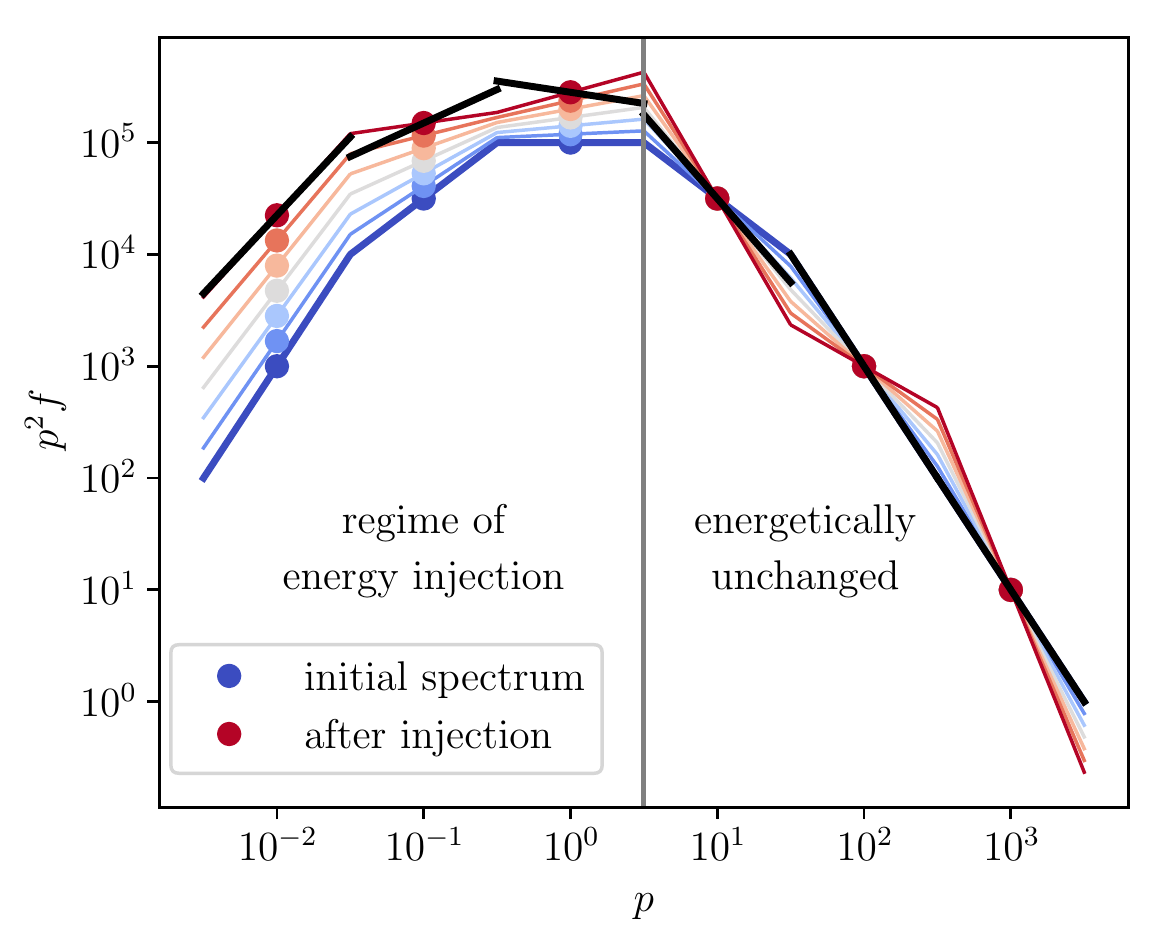}
\caption{Illustration of the problematic treatment of a continuous distribution function. The coloured lines show the evolution of a continuous spectrum during the injection of energy at low momenta. The continuous representation results in local changes in the slope across the entire spectrum. At the high-momentum part, which should effectively be unchanged, the spectrum alternates between a positive and negative curvature spectrum. Avoiding this alternating change in slope in a continuous spectrum after injection would require reshaping the entire spectrum. A discontinuous spectrum can cope with local spectral changes much better (black lines).}
\label{fig:sketch-continuous-problem}
\end{figure}

Several methods have been proposed to solve the Fokker-Planck equation using a piecewise power law representation \citep{JunJones1999,Miniati2001,JonesKang2005,YangRuszkowski2017}. If the numerical scheme is only based on one moment the two degrees of freedom per bin need a closure relation. A simple and intuitive assumption is to require a continuous function for $f$. This reduces the degrees of freedom for $N$ bins to $N+1$, i.e. there is only one additional condition that needs to be set. \citet{JunJones1999} force the slopes in the first two bins to be equal. \citet{Miniati2001} assumes the proton spectrum to be of constant curvature, i.e. $\slope_{i+1}-\slope_i = \slope_i -\slope_{i-1}$. However, there is a fundamental problem with a continuous description of $f$ in particular for a locally varying spectrum. Let us assume a steady state spectrum as shown in the blue curve in Fig~\ref{fig:sketch-continuous-problem}. This spectrum can nicely be represented by a continuous function of piecewise powerlaws. If we now inject energy at the three lowest bins and force the spectrum to be continuous after the injection (dark red curve), we force changes of the local slope across the entire spectrum. The final continuous representation then alternates between a concave and a convex spectrum. Avoiding this alternating behaviour would require to reshape also the high-energy part of the spectrum, which should effectively be unchanged if only low-energy CRs are injected. Without the restriction of a continuous distribution function we can still model the spectrum with a physically useful description (black lines). Even if for most physical applications (injection, cooling, diffusion) there are possibilities to keep a continuous spectrum, the discontinuous representation allows for more freedom and a more stable numerical treatment. The discontinuous modelling, however, requires to constrain two degrees of freedom, which we simply chose to be the two moments of the particle distribution function.

\subsection{Moments of the distribution and their time evolution}

Instead of evolving the momentum-integrated systems in time we need to solve for changes in each momentum bin separately. One special case is advection with the gas, which does not involve spectral changes, but simply advects the spectrum across all bins. We translate the time evolution of the Fokker-Planck equation to the evolution equation for the CR number and energy density in bin $i$. For clarity, we omit the subscript CR in the following two equations. The time evolution of $n_i$ reads
\begin{align}
  \frac{\partial n_{i}}{\partial t} &= \frac{\partial n_{\mathrm{adv},i}}{\partial t} +\frac{\partial n_{\mathrm{diff},i}}{\partial t} + \frac{\partial n_{\mathrm{ad},i}}{\partial t} + \frac{\partial n_{\mathrm{l},i}}{\partial t} + j_{n,i}\\
  &= -\vnabla\bcdot\rkl{\vektor{u}n_{i}} + \vnabla\bcdot\rkl{\skl{\tensor{D}_n}\bcdot\vnabla n_{i}}\notag\\ &+\ekl{\rkl{\frac{1}{3}\rkl{\vnabla\bcdot\vektor{u}}p + b_{\mathrm{l}}(p)} 4\pi\,p^2\,f}_{p_{i-1/2}}^{p_{i+1/2}} + j_{n,i},
\end{align}
where the individual terms describe advection (subscript adv), diffusion (diff), adiabatic changes (ad), and losses (l). Sources are indicated by $j$.
The energy evolution is given by
\begin{align}
  \frac{\partial e_{i}}{\partial t} &= \frac{\partial e_{\mathrm{adv},i}}{\partial t} +\frac{\partial e_{\mathrm{diff},i}}{\partial t} + \frac{\partial e_{\mathrm{ad},i}}{\partial t} + \frac{\partial e_{\mathrm{l},i}}{\partial t} + j_{e,i}\\
  \label{eq:time-evol-energy-density}&= -\vnabla\bcdot\rkl{\vektor{u}e_{\mathrm{cr},i}} + \vnabla\bcdot\rkl{\skl{\tensor{D}_e}\bcdot\vnabla e_{\mathrm{cr},i}}\\
  \label{eq:dedt_ad}&\qquad+ \frac{4\pi}{3}\rkl{\vnabla\bcdot\vektor{u}}\Bigg(\ekl{T(p)p^3f}_{p_{i-1/2}}^{p_{i+1/2}} \Bigg.\\
  &\qquad\qquad\qquad\Bigg.-\int_{p_{i-1/2}}^{p_{i+1/2}} f\frac{p^4c^2}{\sqrt{m^2c^4+p^2c^2}} \dd p\Bigg)\notag\\
  \label{eq:dedt_loss}&\qquad+ 4\pi\ekl{Tp^2b_{\mathrm{l}}(p)f}_{p_{i-1/2}}^{p_{i+1/2}}\\
  &\qquad-4\pi\int_{p_{i-1/2}}^{p_{i+1/2}} \frac{p^3c^2b_{\mathrm{l}}(p)f}{\sqrt{m^2c^4+p^2c^2}} \dd p + j_{e,i}\notag
\end{align}
In both equations (for $n_i$ and $e_i$) we rewrote the diffusion term, such that it formally takes the form of a simple diffusion equation with modified diffusion tensors $\skl{\tensor{D}_n}$ and $\skl{\tensor{D}_e}$,
\begin{align}
  \label{eq:ndiff2}
  \frac{\partial n_{\mathrm{diff},i}}{\partial t} &=\int_{p_{i-1/2}}^{p_{i+1/2}} 4\pi\vnabla\bcdot(\tensor{D}\bcdot\vnabla f)p^2\,\dd p\notag\\
  &= \vnabla\bcdot\rkl{\skl{\tensor{D}_n}\bcdot\vnabla n_{i}}
\end{align}
with
\begin{align}
\label{eq:effective-diff-n}
\skl{\tensor{D}_n} = \rkl{\vnabla n_i}^{-1} \int_{p_{i-1/2}}^{p_{i+1/2}} 4\pi \tensor{D}\,\bcdot\vnabla f\,p^2 \dd p
\end{align}
and analogously for the energy equation,
\begin{align}
\label{eq:effective-diff-e}
\skl{\tensor{D}_e} = \rkl{\vnabla e_i}^{-1} \int_{p_{i-1/2}}^{p_{i+1/2}} 4\pi \tensor{D}\,\bcdot\vnabla f\,p^2T(p)\dd p.
\end{align}
We note that the the inverse of the gradient needs to be computed for every individual component. We discuss the effective diffusion coefficient in more detail in Section~\ref{sec:diffusion}.

\section{Derivation of a numerical scheme for the time evolution}
\label{sec:derivation}

\subsection{Computing changes in number and energy density}

\begin{figure}
\includegraphics[width=8cm]{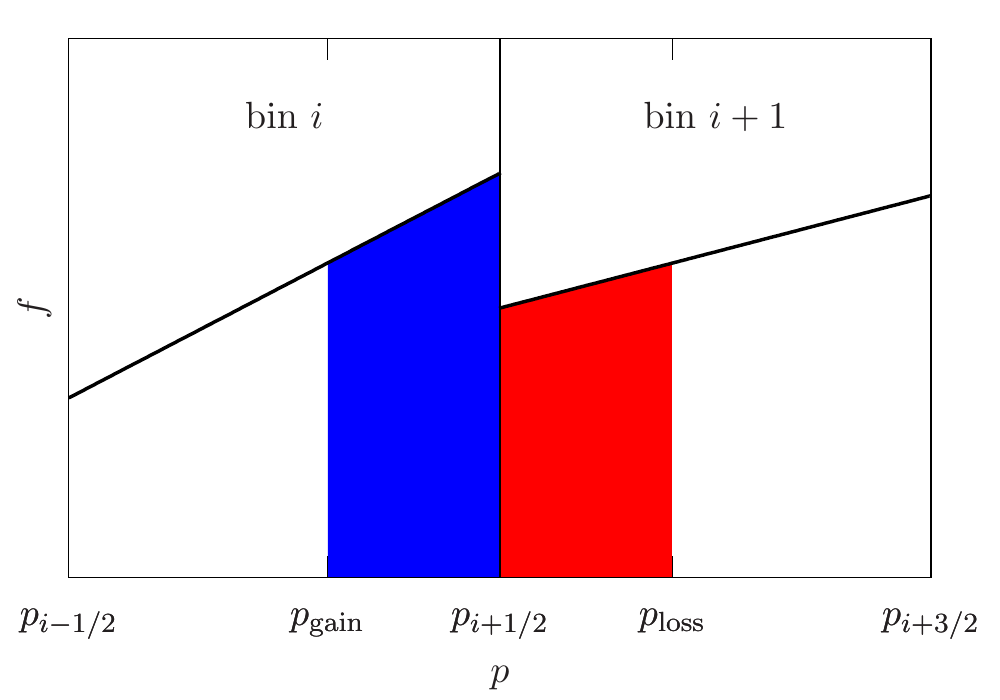}
\caption{Illustration of the spectral shift. A gain in momentum corresponds to a shift of the blue integral across the boundary at $p_{i+1/2}$ while cooling corresponds to shifting the red area to the left.}
\label{fig:sketch-shift-momentum}
\end{figure}

\begin{figure}
\includegraphics[width=8cm]{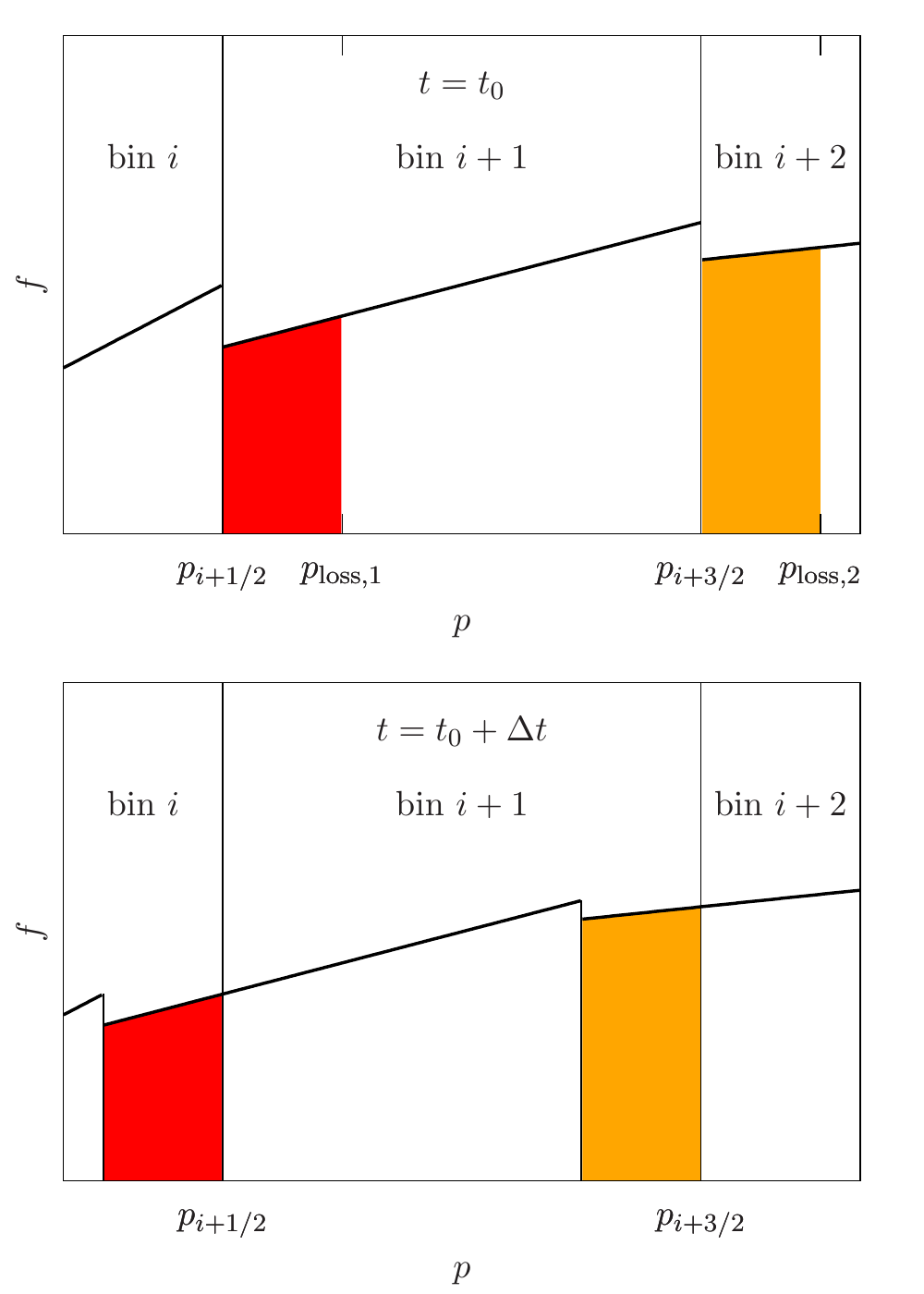}
\caption{Illustration of the time evolution for a loss process before (top) and after the transport step (bottom). The number and energy density corresponding to the red and yellow area are shifted to the lower bins.}
\label{fig:sketch-shift-momentum-time-evol}
\end{figure}

We use operator splitting for the individual parts of the time evolution of the Fokker-Planck equation, in particular the evolution in space and the spectral evolution. In this paper we focus on the spectral evolution and discuss the integration of the method into hydrodynamics in a subsequent paper. We describe the individual parts for the physical processes in terms of the discretised momentum bins with piecewise power laws following \citet{Miniati2001}. In general we convert the evolution in time into an evolution in momentum,
\begin{equation}
\label{eq:gains-losses-general}
\frac{\dd p}{\dd t} = F(p, t, T(p), \vektor{u}),
\end{equation}
where $F$ is a function that depends on the individual physical processes. In particular we cover:
\begin{itemize}
\item adiabatic gains and losses,
\item injection via diffusive shock acceleration,
\item hadronic losses and Coulomb losses, and
\item momentum-dependent diffusion.
\end{itemize}

For any given physical process and a given integration time step $\Delta t$ we can compute the change in momentum. Without loss of generality, we would like to illustrate this for a loss term, $b_\mathrm{l}$,
\begin{equation}
\frac{\dd p}{\dd t} = b_\mathrm{l} < 0,
\end{equation}
which we can rewrite as
\begin{equation}
\label{eq:loss-momentum-time-step}
\int_{p_\mathrm{loss}}^{p_x}\frac{\dd p}{b_\mathrm{l}} = \int_{t_0}^{t_0+\Delta t} \dd t = \Delta t
\end{equation}
if the loss process does not explicitly depend on time. The initial momentum $p_\mathrm{loss}$ cools during $\Delta t$ to $p_x$. For the adiabatic process Eqs.~\eqref{eq:ad-process} and \eqref{eq:adiabatic-compression} explicitly show the closed formulation. We chose $p_x$ to be the momentum at the bin boundary, $p_x = p_{i+1/2}$, and compute the corresponding momentum $p_\mathrm{loss}$, which is illustrated in Fig.~\ref{fig:sketch-shift-momentum}.

At this point we would like to discuss the constraints on the maximum integration time step. As in other explicit numerical methods that account for only the immediate neighbour cells, the flux of number and energy density can extend to at most one bin. For the gain and loss momentum in our setup this means that $p_{i+1/2}\le p_\mathrm{loss} < p_{i+3/2}$ and $p_{i-1/2}< p_\mathrm{gain} \le p_{i+1/2}$. Depending on the loss and gain process, the maximum time step is implicitly given via equation~\eqref{eq:loss-momentum-time-step} by limiting the maximum fraction of the bin that $p_\mathrm{gain}$ or $p_\mathrm{loss}$ should occupy. We find satisfactory results for
\begin{align}
\frac{p_\mathrm{loss}}{p_{i+1/2}} &\le 0.4\left(\frac{p_{i+3/2}}{p_{i+1/2}}\right) \,\mathrm{and}\\
\frac{p_{i+1/2}}{p_\mathrm{gain}} &\le 0.4\left(\frac{p_{i+1/2}}{p_{i-1/2}}\right).
\end{align}

The shift of momentum from $p_\mathrm{loss}$ to $p_{i+1/2}$ corresponds to a transport of particle and energy density across the spectral bin boundary $i+1/2$,
\begin{align}
\Delta n_{i+1/2} &= 4\pi\int_{p_\mathrm{s,gain}}^{p_{i+1/2}} \, p^2 f(p) \dd p,\\
\Delta e'_{i+1/2} &= 4\pi\int_{p_\mathrm{s,gain}}^{p_{i+1/2}} \, p^2 f(p)\,T(p) \dd p,
\end{align}
 illustrated by the red area in Fig.~\ref{fig:sketch-shift-momentum}. This is the essential part of the time evolution of the new method. We evolve the number and energy density in bin $i+1$ as
\begin{align}
n(t+\Delta t) &= n(t) - \Delta n_{i+1/2} + \Delta n_{i+3/2},\\
e'(t+\Delta t) &= e(t) - \Delta e'_{i+1/2} + \Delta e'_{i+3/2}.
\end{align}
The transport of $\Delta n$ from bin $i+1$ to bin $i$ does not require any further correction. Contrary, for the energy term we need to take into account that the loss in momentum connects to a loss in energy, i.e. shifting the spectrum towards lower momenta, we need to additionally correct for that shift,
\begin{equation}
\Delta e_{i+1/2} = \Delta e'_{i+1/2} \,\frac{T(p_{i+1/2})}{T(p_\mathrm{loss,1})}.
\end{equation}
The transfer into and out of cell $i+1$ is illustrated in Fig.~\ref{fig:sketch-shift-momentum-time-evol}. We can compute the energy correction factors independently for $\Delta e$ at interface $p_{i+1/2}$ and $p_{i-1/2}$. In case of identical ratios of $p_\mathrm{loss,1}/p_{i+1/2}$ and $p_\mathrm{loss,2}/p_{1+3/2}$ the energy correction factors for $\Delta e'_{i+1/2}$ and $\Delta e'_{i+3/2}$ are similar. In the non-relativistic limit as well as in the relativistic limit, in which the particle energy has a simple scaling with the momentum ($\propto p^2$ and $\propto p$, respectively), the same ratio for $p_\mathrm{loss,1}/p_{i+1/2}$ and $p_\mathrm{loss,2}/p_{1+3/2}$ results in the same correction factor for the transfer into and out of cell $i+1$, i.e.
\begin{equation}
e(t+\Delta t) = e'(t+\Delta t)\,\frac{T(p_{i+1/2})}{T(p_\mathrm{loss,1})}.
\end{equation}
In the transition region between the classical and the relativistic energy, the factors differ. If the correction factors at the left and right boundary of the bin are not largely different, we can simply apply $\Delta e'_{i+1/2}$ and $\Delta e'_{i+3/2}$ and then correct the entire modified bin $i+1$ by an arithmetic average of the correction factors,
\begin{equation}
e(t+\Delta t) = e'(t+\Delta t)\,\frac{1}{2}\left(\frac{T(p_{i+1/2})}{T(p_\mathrm{loss,1})} + \frac{T(p_{i+3/2})}{T(p_\mathrm{loss,2})}\right).
\end{equation}

Depending on the process the ratios $p_\mathrm{loss,1}/p_{i+1/2}$ and $p_\mathrm{loss,2}/p_{i+3/2}$ behave differently. For adiabatic expansion (and compression) both ratios are the same as well as for the hadronic losses as long as all momenta are above the threshold momentum for hadronic losses. In these cases, the energy correction factor for the bin is trivial. The case of Coulomb losses at sub-relativistic energies is more difficult to compute. As the loss rate scales with approximately $p^{-1.9}$, the loss rate at the lower momentum boundary can be significantly larger than at the high momentum boundary, in particular, if the bins span half an order of magnitude. In this regime we therefore compute the correction term numerically using a subgrid interpolation. We subdivide the bin in 10 logarithmically spaced sub-bins with momenta $p_j(t_0)$ and compute the momenta after $\Delta t$, $p_j'(t_0=\Delta t)$, using Eqs.~\eqref{eq:gains-losses-general} and \eqref{eq:coulomb-loss-approximate}. We then average over the individual values $\langle p_j'(t_0+\Delta t)\rangle_j$.

In principle, we can compute a total shift for each bin for all terms in the Fokker-Planck equation. However, it is simpler to compute a separate shift momentum for the individual processes. The adiabatic process can also be used to analytically show that the method of discretisation yields correct results, which we demonstrate in appendix~\ref{sec:adiabatic-analytic-proof}.

\subsection{Reconstruction of the particle distribution function}
\begin{figure}
\includegraphics[width=8cm]{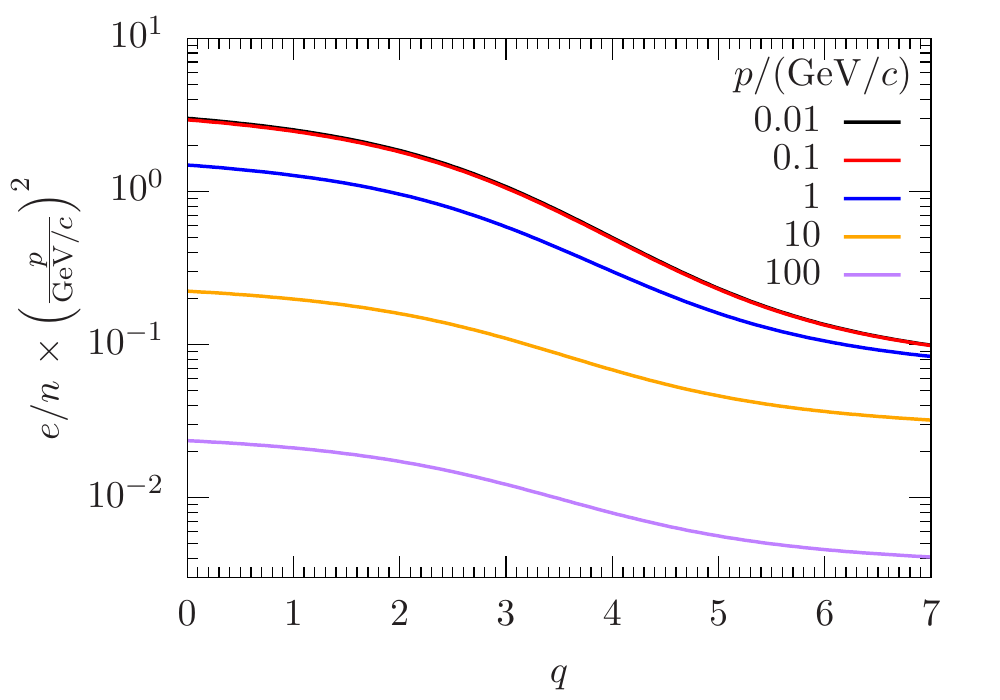}
\caption{Scaling of $e/n$ as a function of spectral slope $\slope$. We multiply the curves by $p^2$ for better illustration. For all relevant slopes there is a one-to-one mapping between $e/n$ and $\slope$. This allows for an exact reconstruction of $f_{i-1/2}$ and $\slope_i$ after a change of $n$ and $e$.}
\label{fig:scaling-en-ratio-slope}
\end{figure}

After computing the temporal changes we end up with a modified number and energy density in each bin. We now need to reconstruct the new amplitude and slope of the particle distribution function in every bin. We first compute the new slope by solving the ratio
\begin{align}
\frac{e_i}{n_i} &= \frac{\int_{p_{i-1/2}}^{p_{i+1/2}}4\pi p^2 f(p) T(p)\,\dd p}{\int_{p_{i-1/2}}^{p_{i+1/2}}4\pi p^2 f(p) \dd p}\notag\\
&= \frac{\int_{p_{i-1/2}}^{p_{i+1/2}}4\pi p^2 f_{i-1/2}\,(p/p_{i-1/2})^{-\slope_i} \, T(p)\,\dd p}{\int_{p_{i-1/2}}^{p_{i+1/2}}4\pi p^2 f_{i-1/2}\,(p/p_{i-1/2})^{-\slope_i}\,\dd p}
\end{align}
for the slope $\slope_i$ numerically using the Newton-Raphson method. The solution of $\slope_i$ is unique as long as number density and energy density scale differently as a function of the slope. In Fig.~\ref{fig:scaling-en-ratio-slope} we plot the ratio $e/n$ as a function of $\slope$ for different energies. For a physically relevant range of slopes there is a one-to-one mapping allowing a unique reconstruction. We tabulate values at the beginning of the simulation to speed up the computation. As the integral for the number density has a simple closed form, we can trivially solve for the amplitude $f_{i-1/2}$ analytically and compute it using the new slope $\slope_i$. We could also use the integral for the energy to find $f_{i-1/2}$, however, the integral in general cannot simply be solved for $f_{i-1/2}$ analytically.

\subsection{Spectral boundary conditions}

In order to conserve CR energy, one would need to apply closed boundaries. In case of losses the energy would accumulate in the lower buffer bin. Adiabatic gains due to strong compression would be stored in the upper buffer bin. Although this seems like a reasonable process for one step it bears difficulties over a longer simulation time. Let us assume a long-term cooling period followed by a strong compression as a simple gedanken experiment. The cooling period will result in an accumulation of significant CR energy in the lowest bin with a large amplitude $f_0$. Even if the CRs cooled entirely to $p=0$ (see Fig.~\ref{fig:cooling-times}) over that period, they would still be buffered in the lower buffer bin with effectively non-negligible momentum. The strong compression will then push this high-amplitude bin to larger momentum bins resulting in an artificially large population of CRs in bins $i$ that are affected by the compression, cf. equation~\eqref{eq:adiabatic-compression}.

We therefore effectively use different boundary conditions for the individual processes. We note that hadronic losses only occur for CR momenta above $p_\mathrm{thr}$, so for spectral configurations with $p_\mathrm{min}<p_\mathrm{thr}$ the hadronic losses do not interfere with the lower boundary. At the high-$p$ boundary we acknowledge that our numerical spectrum is only a small cut-out of the total CR spectrum, which extends as a power-law tail up to CR momenta of $p\sim10^{20}\,\mathrm{eV}/c$. We therefore use inflow boundary conditions, which reflect the continued spectrum towards higher momenta. In practice we simply keep the slope at the highest bin constant during the hadronic cooling step, i.e. it is determined by a combination of the injection spectrum from the previous time step and by the slope from the next lower momentum bin in case of adiabatic compression. We note that energy dependent spatial diffusion would also change the slope inside every bin. However, we currently only account for energy dependent diffusion of each independently, see Section~\ref{sec:diffusion}. For Coulomb losses, we apply outflow boundary conditions at small momenta, resulting in a loss of energy for CRs that cool below $p_\mathrm{min}$, which mimics the
thermalisation process of these CRs. At the high-momentum boundary we follow the same reasoning as for the hadronic losses and compute an inflow of energy based on the continued spectrum outside of our spectral range. For the adiabatic process we assume that the slope in the inflowing buffer bin (lower buffer bin for compression and vice versa) does not change and we allow CRs to enter and leave the spectrum during compression and expansion.

\section{One-zone tests}
\label{sec:one-zone}

\begin{figure*}
\begin{minipage}{\textwidth}
\includegraphics[width=0.5\textwidth]{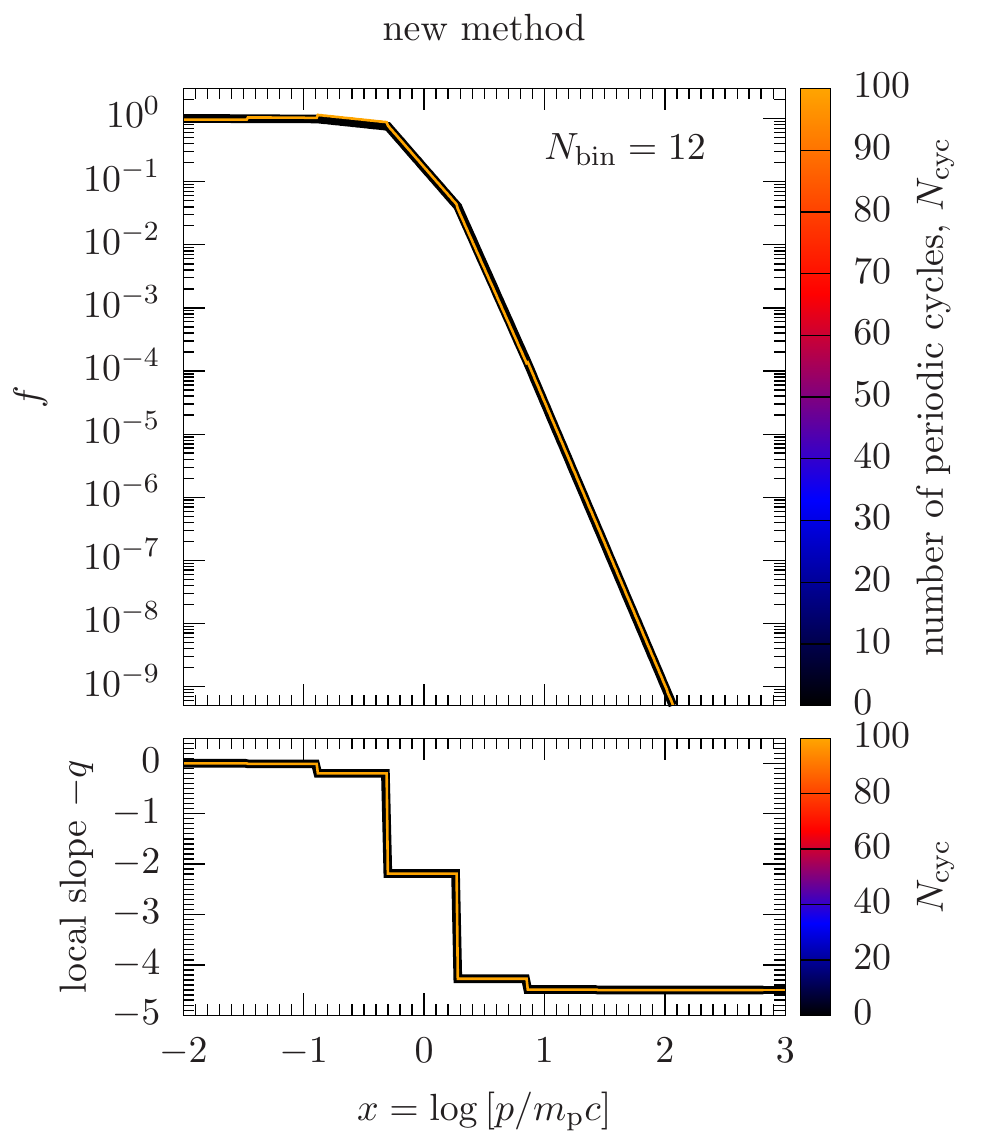}
\includegraphics[width=0.5\textwidth]{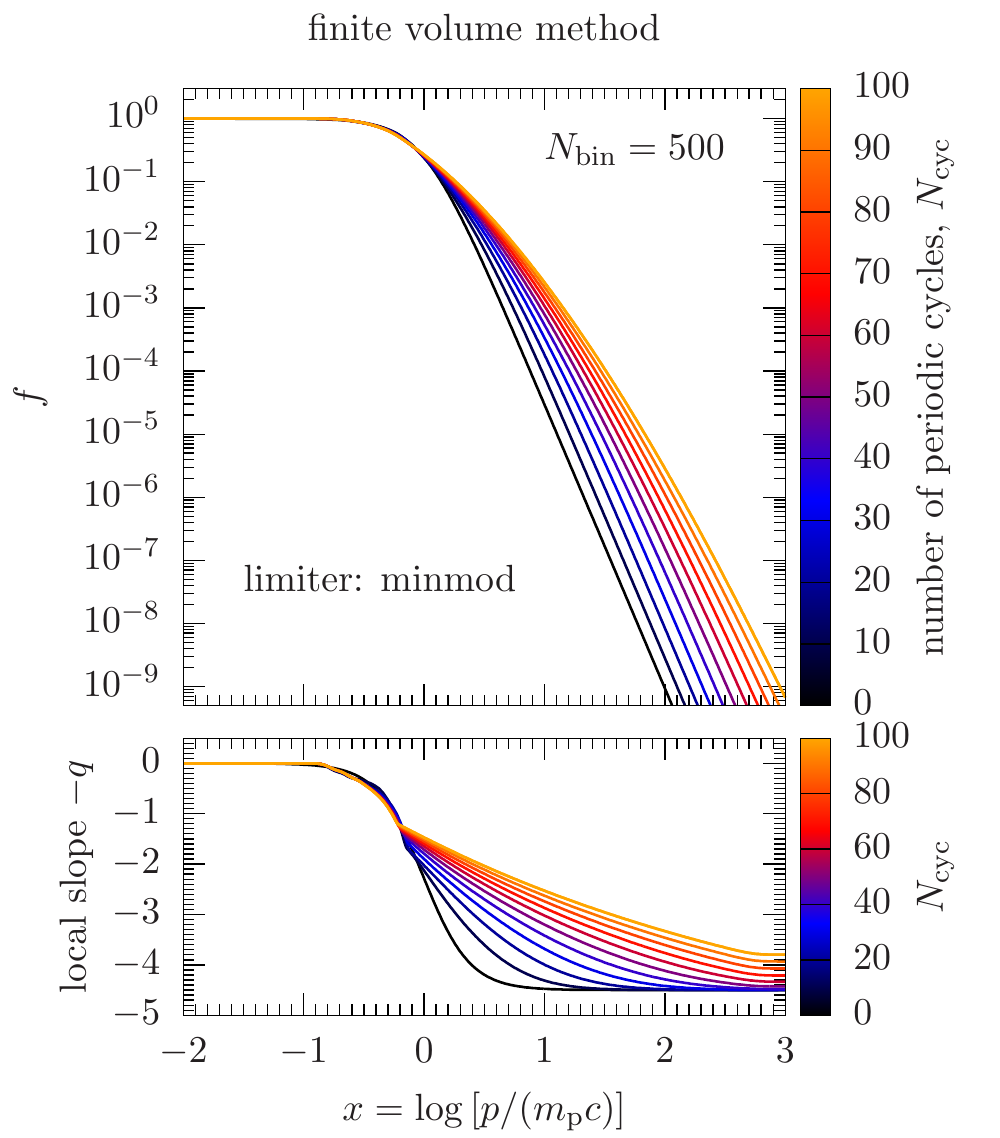}
\caption{Numerical solution of the spectrum for periodic adiabatic compression and expansion cycles with a momentum compression ratio of $\approx2.2$ (density compression ratio of 10) using the initial spectrum described in equation~\eqref{eq:analytic-2-pl-spectrum}. Shown are the spectra after every 10$^\mathrm{th}$ full cycle with the time colour coded. The left-hand panel shows the new method with piecewise powerlaws for the particle distribution function using 12 spectral bins for 5 orders of magnitude in momentum. The right-hand panel shows the piecewise constant counterpart with a finite volume method and piecewise constant representation of $f$ using 500 bins. The new method accurately restores the initial spectrum after every cycle. Numerical diffusion results in a broadened spectrum over time using the conventional method.}
\label{fig:adiabatic-losses-comparison}
\end{minipage}
\end{figure*}

\begin{figure}
\includegraphics[width=0.5\textwidth]{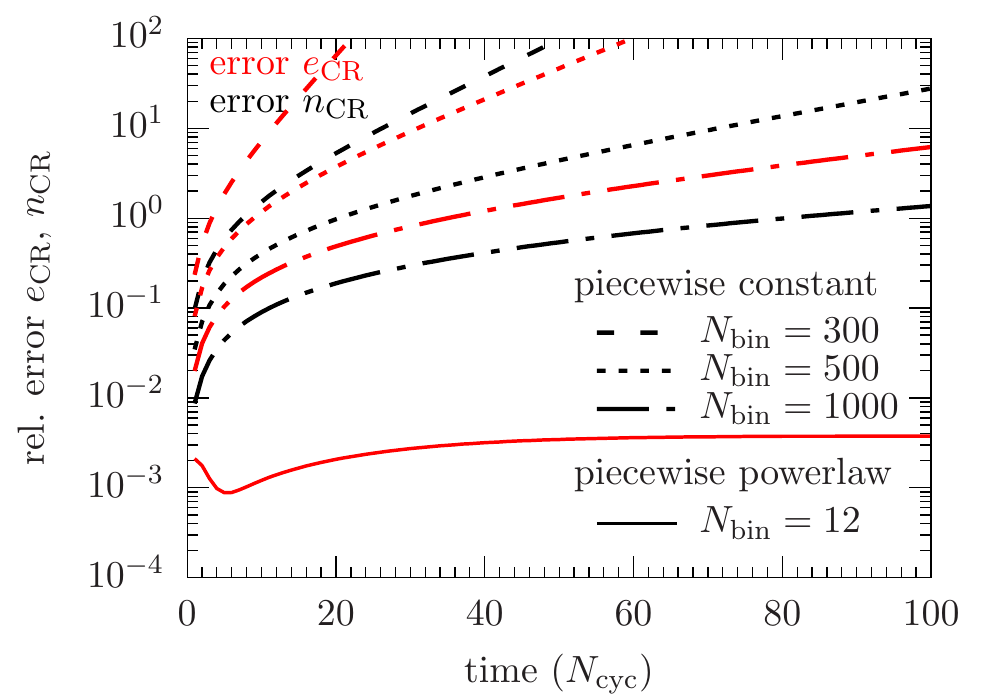}
\caption{Relative error of the number and energy density for a piecewise constant representation of $f$ (dashed lines) and our new scheme (solid lines) for the spectral setups shown in Figs.~\ref{fig:adiabatic-losses-comparison}. The error is shown as a function of time in units of periodic cycles with a momentum compression factor of $\approx2.2$. For the new method the error in the number density is again several orders of magnitude smaller (not shown). The conventional method using 1000 spectral bins still shows an error which is 3 orders of magnitude larger than the computation with the new method and only 12 bins.}
\label{fig:adiabatic-losses-error}
\end{figure}

\begin{figure}
\includegraphics[width=8cm]{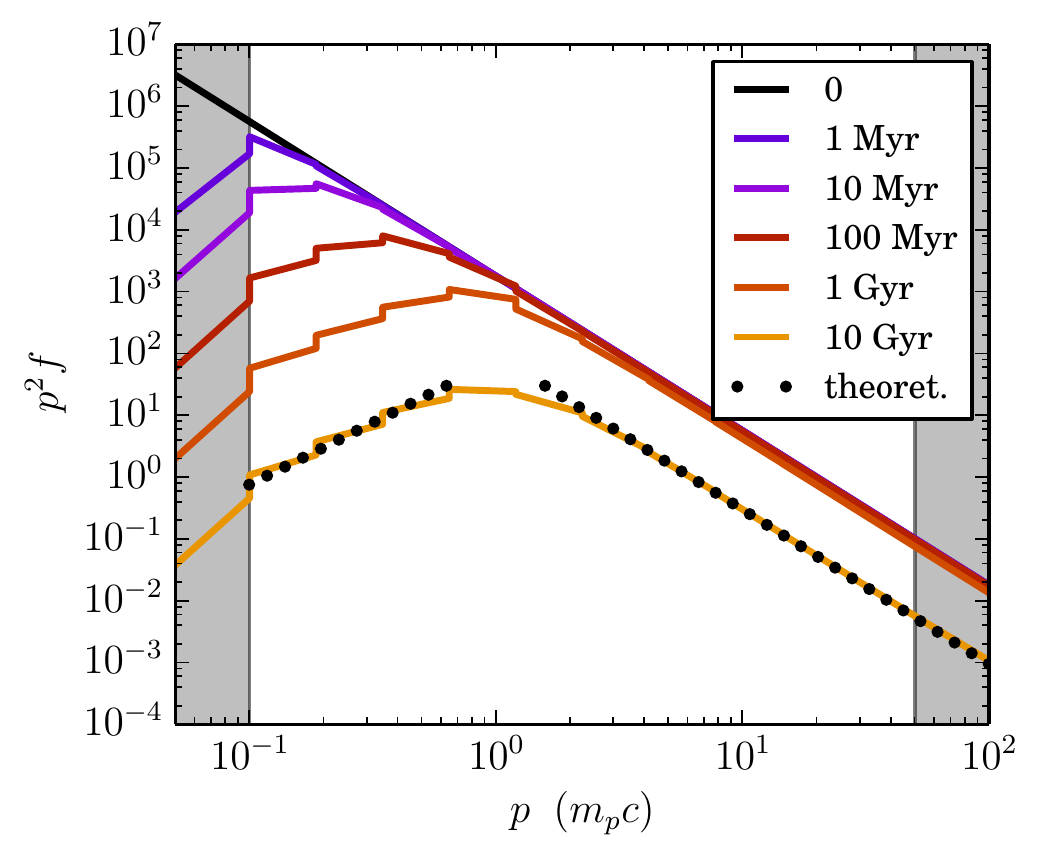}
\caption{Time evolution of the freely cooling spectrum performed with 10 spectral bins. The two dotted lines represent the asymptotic limits of the freely cooling CR distribution, which decays in amplitude over time. We adopt $n_\mathrm{e}=n_\mathrm{N}=10^{-2}\,\mathrm{cm}^{-3}$.}
\label{fig:free-cooling-time-evol}
\end{figure}

\begin{figure}
\includegraphics[width=8cm]{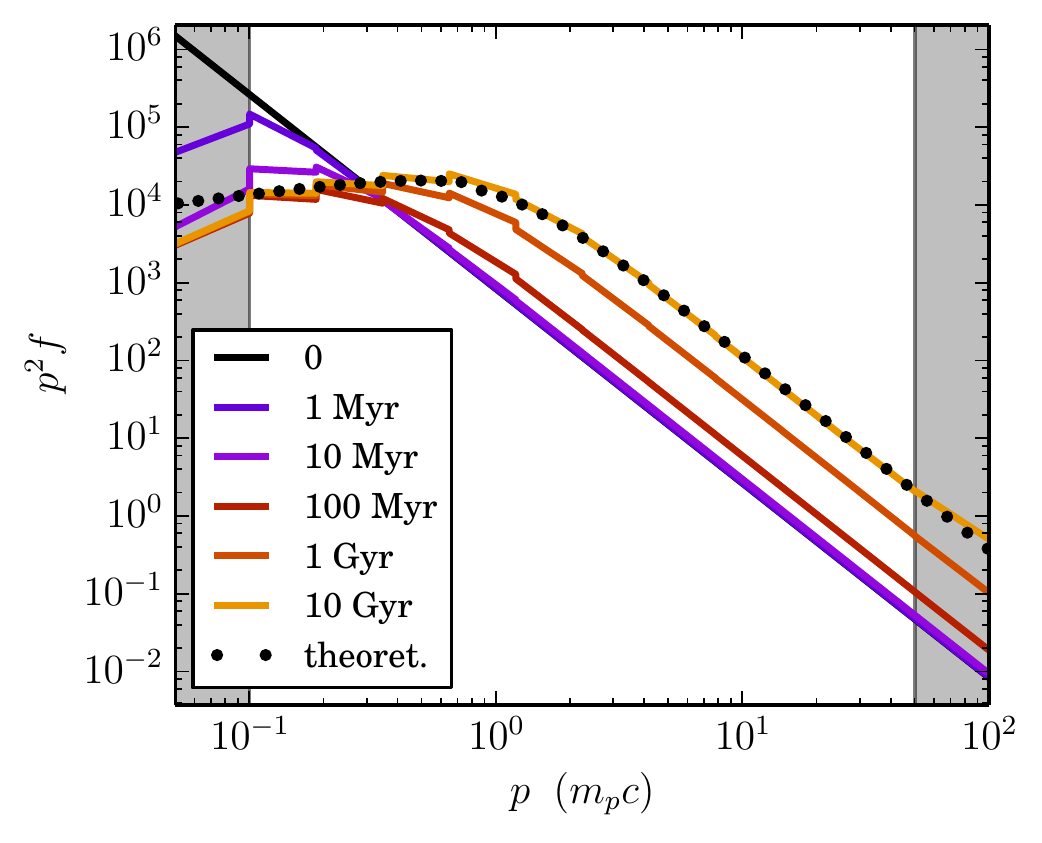}\\
\includegraphics[width=8cm]{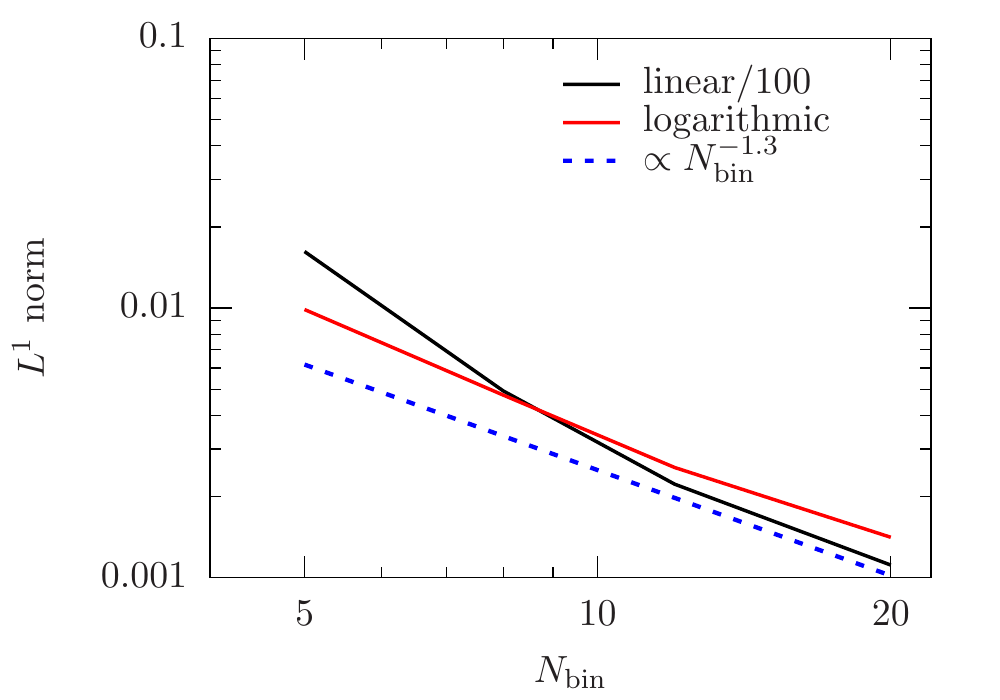}
\caption{\textsc{Top}: Time evolution of the spectrum following continuous injection and cooling. The initial spectrum increases at high energies and cools at low energies to reach the steady state after 10 Gyr. The computation is performed with 10 spectral bins. The theoretical curve derives from equation~\eqref{eq:steady-state}. We adopt $n_\mathrm{e}=n_\mathrm{N}=10^{-2}\,\mathrm{cm}^{-3}$. \textsc{Bottom}: Error of the numerical method compared to the analytic spectrum encoded in the $L^1$ norm as a function the number of spectral bins, $N_\mathrm{bin}$. The error scales approximately as $N_\mathrm{bin}^{-1.3}$.}
\label{fig:ss-time-evol}
\end{figure}

\subsection{Test of the adiabatic process}

We would like to stress the importance of an accurate computation of the adiabatic process in hydrodynamical simulations. In compressive turbulent environments the gas cells frequently experience changes of short compression and expansion periods. Even without fully developed turbulence, pressure waves travelling through the simulation domain result in numerous small oscillations.

For this test we only consider adiabatic compression and expansion. Injection of CRs as well as other loss processes are switched off. We compare our new method with a standard finite volume method \citep[e.g.][]{Toro2009} for the piecewise constant representation of the particle distribution function.

As initial condition we use a simple analytic function that mimics a steady state spectrum, i.e. a flat slope at low momenta and a scaling of $f\propto p^{-4.5}$ at high momenta,
\begin{equation}
\label{eq:analytic-2-pl-spectrum}
f(t=0) = A_0\,\left[\left(\frac{p}{\mathrm{1GeV/}c}\right)^{-a/c} + \left(\frac{p}{\mathrm{1GeV/}c}\right)^{-b/c}\right]^{-c}.
\end{equation}
The parameter $a$ is the approximate slope for low momenta ($p\rightarrow0$), which we set to $a=0$. Parameter $b$ is the high-momentum counterpart with a value of $-4.5$ and $c\equiv2$ determines the width of the transition region between the two powerlaw regimes. The overall amplitude of the spectrum, $A_0$, is set to unity.

Motivated by the strong dynamics in many astrophysical systems, we do not test the methods only for one compression or expansion step but rather for hundreds of periodic oscillations. The divergence of the velocity $\vnabla\bcdot\vektor{u}$ is modelled as
\begin{equation}
\label{eq:periodic-oscillations}
(\vnabla\bcdot\vektor{u})(t) = A_\mathrm{per}\cos\left(\frac{2\pi t}{T}\right).
\end{equation}
We set $A_\mathrm{per}=7.35$ and $T=1.0$. Following equation~\eqref{eq:adiabatic-compression-0} we find
\begin{equation}
\frac{\dd p}{p} = -\frac{A_\mathrm{per}}{3}\cos\left(\frac{2\pi t}{T}\right) \dd t,
\end{equation}
which yields
\begin{equation}
p(t) = p(t_0) \exp\left[-\frac{A_\mathrm{per}T}{6\pi}\left\{\sin\left(\frac{2\pi t}{T}\right)-\sin\left(\frac{2\pi t_0}{T}\right)\right\}\right],
\end{equation}
where we set $t_0=0$. After the first quarter of the period, $t=0.25$, the momentum reduces to $68\%$ of the initial value. After three quarters of the period the maximum momentum is reached with an increase to $148\%$ with respect to the initial value. Over one period the ratio of maximum to minimum momentum $p_\mathrm{max}/p_\mathrm{min}\approx2.2$, which corresponds to a density compression factor of $(p_\mathrm{max}/p_\mathrm{min})^3\approx10$. We note that after a full compression and expansion cycle the spectrum needs to take again the initial shape.

Figure~\ref{fig:adiabatic-losses-comparison} shows the spectral evolution over 100 periodic cycles for the new method (left-hand panel) and a classical finite-volume method (right-hand panel) using a minmod limiter. The top panel shows the spectrum, the lower one indicates the slope. Colour-coded is the time in units of full cycles. For the classical method we plot every 10$^\mathrm{th}$ cycle. In the case of the new method the deviations are so small that we only plot the initial function with a thicker black line and the spectrum after 100 cycles. For the new method we use 12 spectral bins, for the conventional finite volume method we show an example with 500 bins. The flat part of the spectrum is accurately solved with the conventional finite volume method. However, at the transition to the high-momentum powerlaw numerical diffusion causes the spectrum to broaden towards higher momentum with an inaccurate representation of the slope.

In a more quantitative way, we investigate the relative error of the number and energy density as a function of periodic cycles, $N_\mathrm{cyc}$, with respect to the initial value,
\begin{equation}
\delta_e = \left|\frac{e_\mathrm{CR}(N_\mathrm{cyc})-e_\mathrm{CR}(0)}{e_\mathrm{CR}(0)}\right|,
\end{equation}
which is plotted in Fig.~\ref{fig:adiabatic-losses-error}. We vary the number of bins from 300 to 1000 for the piecewise constant representation of $f$. In order to reach a relative error of order unity the conventional method needs more than 1000 bins for a momentum ranging over 5 orders of magnitude. The relative error in energy for the new method is of order $10^{-3}$ for only 12 bins, which demonstrates the superior performance of the new scheme. The error in number density is again several orders of magnitude lower than the energy error and is not shown.

Hadronic and Coulomb losses are best tested in a freely cooling test and a steady state solution. In the case of free cooling we start the simulation with a powerlaw $f(p)=10^3(p/(m_\mathrm{p}c))^{-4.5}$ and let the spectrum cool including hadronic and Coulomb losses. We compute the spectrum using sub-cycling. Figure~\ref{fig:free-cooling-time-evol} shows the freely cooling spectrum for different times using 12 spectral bins including buffer bins. Also shown are the approximate slopes at large cooling times indicating the accuracy of the method. As expected (Section~\ref{sec:free-cooling}) in the case of hadronic cooling the cooled spectrum scales as the original spectrum. The Coulomb losses approach a flat spectrum for $f$, i.e. a spectral scaling with $p^{1.9}$ using the scaling in the plot. There is no analytic solution for the amplitude of the spectrum, so we need to focus on the accuracy of the slopes.

For the steady state spectrum we use again the same initial spectrum as in the case for free cooling, $f(p)=10^3(p/(m_\mathrm{p}c))^{-4.5}$. We then apply a constant injection $j(p)=2.58/\mathrm{Myr}(p/(m_\mathrm{p}c))^{-4.5}$, i.e. a fraction of $2.58\times10^{-3}$ of the energy of the initial spectrum. We use time steps of $\Delta t =2\,\mathrm{Myr}$ and in each time step we apply injection, Coulomb and hadronic losses using 12 bins including buffer bins. The spectrum is shown in the top panel of Fig.~\ref{fig:ss-time-evol} for different times. We overplot the numerical solution at $10\,\mathrm{Gyr}$ with the analytical steady state spectrum as in equation~\eqref{eq:steady-state}. The converged spectrum agrees well with the analytic solution. More quantitatively, the bottom panel shows the error of the numerical solution in comparison to the analytic result using the $L^1$ norm. We adopt a linear measure,
\begin{equation}
L^1_\mathrm{lin} = N^{-1} \sum_i |f_{i,\mathrm{num}}(p_i)-f_{i,\mathrm{ana}}(p_i)|
\end{equation}
as well as its logarithmic counterpart,
\begin{equation}
L^1_\mathrm{log} = N^{-1} \sum_i |\log (f_{i,\mathrm{num}}(p_i)/f_{i,\mathrm{ana}}(p_i))|,
\end{equation}
where in both cases we sample the spectrum with 250 data points per momentum decade, giving rise to $N=624$ sampling points. The error reduces with a scaling of approximately $L^1\propto N_\mathrm{bin}^{-1.3}$.

\section{Spatial diffusion}
\label{sec:diffusion}

So far we have discussed the spectral evolution without investigating the connection to the spatial time evolution. For advection, this is only a problem of the hydrodynamical code, which is beyond the scope of this paper. Spatial diffusion in contrast reveals an interaction of spatial and spectral changes. As spatial diffusion is energy conserving, the diffusion step itself does not transfer CRs in momentum space, i.e., $\partial f/\partial p=0$. However, the amount of diffused number and energy density to neighbouring hydrodynamical cells will depend on the spatial derivatives of $f$ and the possible different diffusion speeds for $n$ and $e$ will result in changes of $f_{i-1/2}$ and $\slope_i$ within one bin. This in turn will affect the other processes in the following time step. We have to look at diffusion separately for number density and energy density.

\subsection{Diffusion of number density}

The spatial diffusion term of CRs in the momentum range $[p_1, p_2]$ is
\begin{align}
  \label{eq:ndiff}
  \partial_t n_\mathrm{diff} &= \int_{p_1}^{p_2} 4\pi\vnabla\bcdot(\tensor{D}\bcdot\vnabla f)p^2\,\dd p\notag\\
  &=4\pi\vnabla \bcdot\ekl{\int_{p_1}^{p_2} p^2 \tensor{D}\bcdot\vnabla f\,\dd p},
\end{align}
where $\tensor{D}$ is the spatial CR diffusion tensor,
\begin{align}
  \tensor{D} &= \rkl{\begin{array}{ccc}
      D_{11} & D_{12} & D_{13}\\
      D_{21} & D_{22} & D_{23}\\
      D_{31} & D_{32} & D_{33}
  \end{array}}
\end{align}
whose components reflect the orientation of the magnetic fields. Following \citet{RyuEtAl2003}, we set the components
\begin{equation}
  D_{ij} = D_{\perp}\delta_{ij} + (D_{\parallel} - D_{\perp})b_ib_j
\end{equation}
with the normalised magnetic field components $b_i=B_i/|\vektor{B}|$. The diffusion parameters depend on momentum, so
\begin{align}
\label{eq:diff-coeff-scaling}
  D_{\parallel}(p) &= D_{\parallel,10} \rkl{\frac{p}{p_{10}}}^{\alpha}\\
  D_{\perp}(p) &= D_{\perp,10} \rkl{\frac{p}{p_{10}}}^{\alpha},
\end{align}
where $p_{10} = 10\,\mathrm{GeV/c}$ and $D_{\parallel,10}$ and $D_{\perp,10}$ are diffusion coefficients parallel and perpendicular to the magnetic field line for momentum $p_{10}$. We can solve equation~\eqref{eq:ndiff} directly by replacing the individual components, which results in 
\begin{align}
  &\partial_t n_\mathrm{diff} =\notag\\
  &4\pi\,\vnabla\bcdot\,\int_{p_1}^{p_2}\,p^2\,\ekl{\begin{array}{c}
      D_{11}\partial_xf + D_{12}\partial_yf + D_{13}\partial_zf\\
      D_{21}\partial_xf + D_{22}\partial_yf + D_{23}\partial_zf\\
      D_{31}\partial_xf + D_{32}\partial_yf + D_{33}\partial_zf
  \end{array}}\,\dd p,
\end{align}
where $\partial_k=\partial/\partial k$ is a shorthand notation for the partial derivative with $k\in\{x, y, z\}$. In order to solve this equation we need to compute the spatial derivatives of $f$. Alternatively, we can write the diffusion equation as
\begin{align}
  \label{eq:ndiff_alt}
  \frac{\partial n_\mathrm{diff}}{\partial t} &=\int_{p_1}^{p_2} 4\pi\vnabla\bcdot(\tensor{D}_n\bcdot\vnabla f)p^2\,\dd p\notag\\
  &= \vnabla\bcdot\rkl{\skl{\tensor{D}_n}\bcdot\vnabla n_{cr}},
\end{align}
such that it formally takes the form of a simple diffusion equation with modified diffusion tensor $\skl{\tensor{D}_n}$, see equations~\eqref{eq:effective-diff-n}.
We arrive at that equation in a simple intuitive way formally by multiplying with unity. The individual tensor components take the form
\begin{equation}
  \skl{D_{n,ij}} = \frac{\int_{p_1}^{p_2} p^2\,D_{ij}\,\partial_j f\,\dd p}{\int_{p_1}^{p_2} p^2 \partial_j f\,\dd p}
\end{equation}
Using $D_{ij}=D'_{ij}(p/p_{10})^\alpha$ we get
\begin{equation}
  \skl{D_{n,ij}} = \frac{D'_{ij}}{p_{10}^\alpha}\,\frac{\int_{p_1}^{p_2} p^{2+\alpha}\partial_j f\,\dd p}{\int_{p_1}^{p_2} p^2 \partial_j f\,\dd p}
\end{equation}

\subsection{Diffusion of energy density}

In a similar way the diffusion of CR energy density obeys
\begin{align}
  \label{eq:ediff}
  \partial_t e_\mathrm{diff} &= \int_{p_1}^{p_2} 4\pi\vnabla\bcdot(\tensor{D}\bcdot\vnabla f)p^2T(p)\,\dd p\notag\\
  &=4\pi\vnabla\bcdot \ekl{\int_{p_1}^{p_2} p^2T(p) \tensor{D}\bcdot\vnabla f\,\dd p},
\end{align}
Using equation~\eqref{eq:effective-diff-e} we obtain modified diffusion coefficients for the energy
\begin{equation}
\skl{D_{e,ij}} = \frac{\int_{p_1}^{p_2} p^2T(p)\,D_{ij}\,\partial_j f\,\dd p}{\int_{p_1}^{p_2} p^2T(p) \partial_j f\,\dd p},
\end{equation}
which can be connected to the scaling of the diffusion coefficients with the momentum to yield
\begin{equation}
\skl{D_{e,ij}} = \frac{D'_{ij}}{p_{10}^\alpha}\,\frac{\int_{p_1}^{p_2} p^{2+\alpha}T(p)\partial_j f\,\dd p}{\int_{p_1}^{p_2} p^2T(p) \partial_j f\,\dd p}.
\end{equation}
\subsection{Simplified bin-centred diffusion}

\begin{figure*}
\begin{minipage}{\textwidth}
\includegraphics[width=\textwidth]{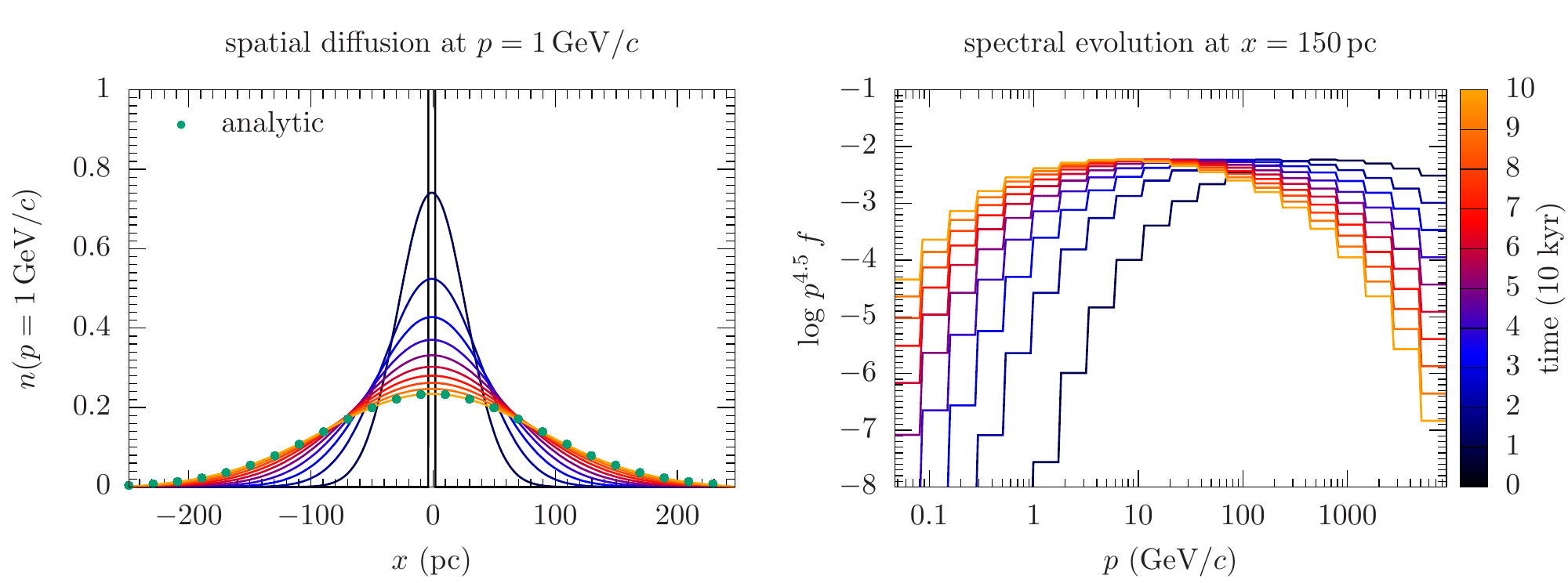}
\caption{Spatial diffusion in one dimension. On the left-hand side we show the diffusion of $n$ of the $1\,\mathrm{GeV}/c$ bin over time including the analytical solution of the last time step. The right-hand side illustrates the spectrum at $150\,\mathrm{pc}$ over time. The initial spectrum is a power-law, $f\propto p^{-4.5}$. At early times the spectrum is dominated by high-energy CRs because they diffuse faster and reach this point first. At later times the low-energy CRs catch up. At this time the amplitude of the highest energy CRs has already decreased due to further diffusion. Locally, the initial slope is conserved in our simplified approximation.}
\label{fig:spatial-diffusion-x}
\end{minipage}
\end{figure*}

\begin{figure*}
\begin{minipage}{\textwidth}
\includegraphics[width=0.48\textwidth]{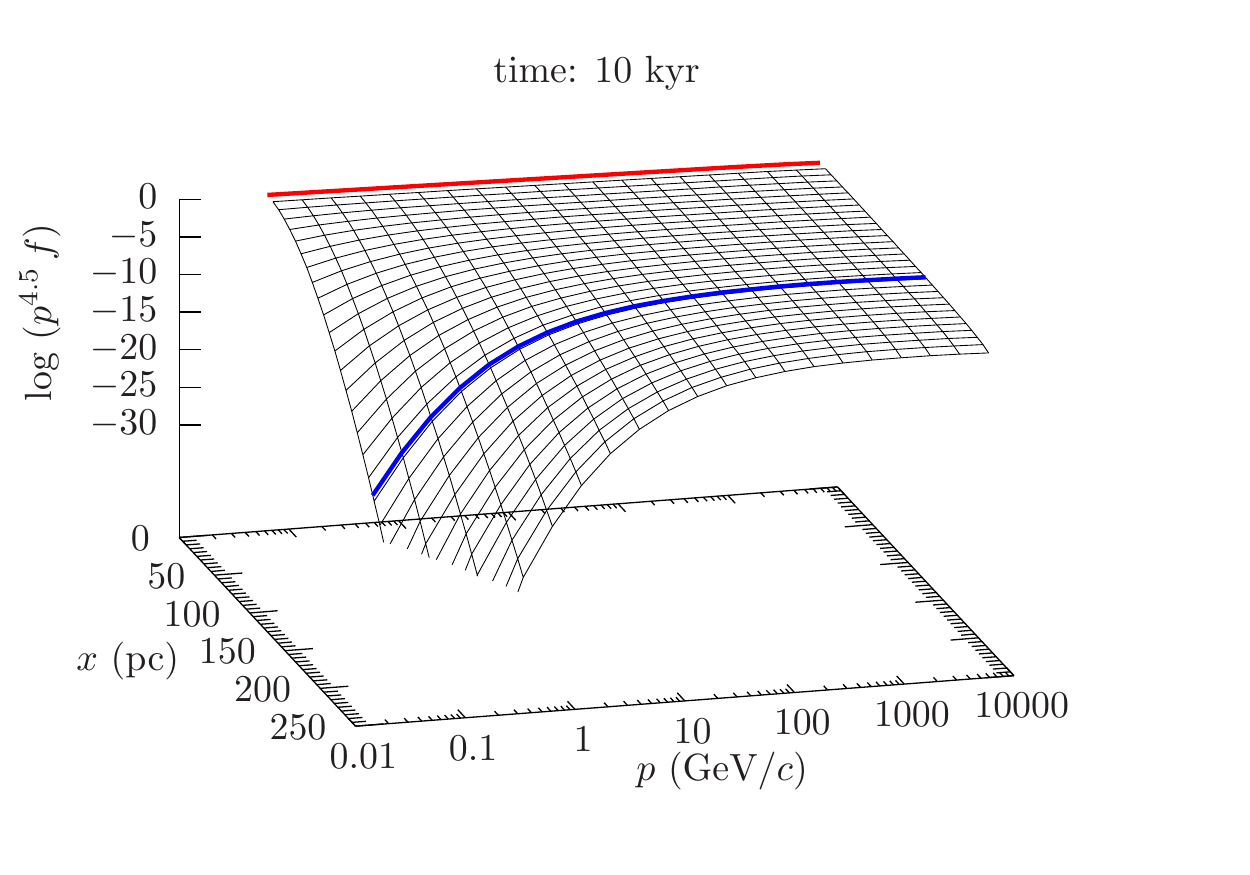}
\includegraphics[width=0.48\textwidth]{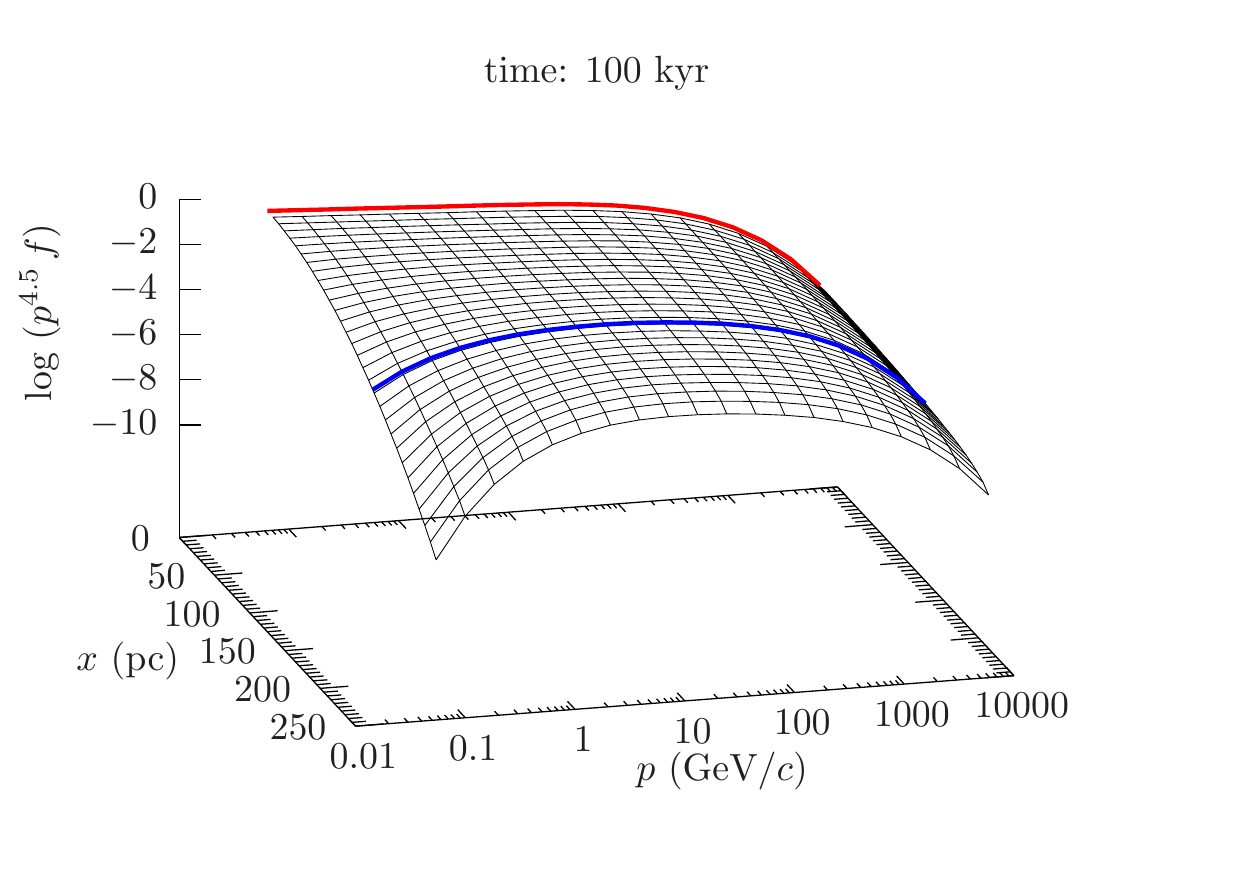}
\caption{Three-dimensional view of the amplitude of the spectrum as a function of momentum and spatial distribution. Shown is only half of the distribution ($x>0$) after 10 (left) and 100 kyr (right). The red (blue) line indicates the spectrum at $x=0$ ($x=200\,\mathrm{pc}$). At early times the low-energy CRs did not have enough time to diffuse to large spatial distances resulting in a steep decline of the blue curve at low $p$. After $100\,\mathrm{kyr}$ low-energy CRs diffused to $x=200\,\mathrm{pc}$ (we note the much smaller dynamical range in $z$). A non-negligible fraction of high-energy CRs started to diffuse out of the box, which leads to a drop both at $x=0$ and $x=200\,\mathrm{pc}$.}
\label{fig:spatial-diffusion-3D}
\end{minipage}
\end{figure*}

In principle, the correct diffusion can be computed using the previously derived coefficients. However, there are two disadvantages of the computations. One is the computational cost of the spatial derivative because it includes both the spatial derivative of the amplitude $f_{i-1/2}$ as well as the slope $\slope_i$. This expensive derivative has to be done for every cell and every momentum bin separately, which means that no part of the diffusion tensor can be reused for several spectral bins. A second complication is that the full spatial derivative of $f$ has a large dynamic range, in particular near CR sources where the spectrum can differ by orders of magnitude. The strong anisotropy with approximately two orders of magnitude larger parallel than perpendicular diffusion further increases the numerical demands. The derivatives and a stable numerical diffusion are therefore difficult to control in this case. For conserved quantities, slope limiters could be used for the fluxes.

Instead of computing the spatial derivatives of $f$ we can use a simplified approximation and consider the diffusion of number and energy density with the same momentum-dependent diffusion coefficients evaluated with diffusion coefficients for the bin centres of the momenta $p_i$. This equal treatment of $n$ and $e$ will result in a non-varying slope $\slope_i$ for the diffusion step. This is an approximation that is reasonably stable even in the case of strong shocks and strong CR injection combined with highly anisotropic diffusion.

A few conceptual properties of spatial momentum dependent diffusion can be illustrated in a one-dimensional toy model. We set up an initial Dirac $\delta$ distribution in CR number and energy density. We rescale the amplitude in every momentum bin to give a spectral power-law with index $-4.5$ corresponding to injection of shock-accelerated CRs with a shock of moderate strength. The CR are allowed to diffuse spatially as indicated in the left-hand panel of Fig.~\ref{fig:spatial-diffusion-x} for the spectral bin at $p=1\,\mathrm{GeV}/c$, where we applied an arbitrary renormalisation for better readability and use a diffusion coefficient of $D_{\parallel,10} = 10^{28}\,\mathrm{cm}^2\,\mathrm{s}^{-1}$ and a scaling of $\alpha=0.5$, cf equation~\eqref{eq:diff-coeff-scaling}. With this approximation all momentum bins follow the same spatial functional evolution over time, which is a Gaussian function. However, the different diffusion speeds result in faster diffusion for higher CR momenta. The initial power-law spectrum therefore changes its shape over time, which is shown in the right-hand panel of Fig.~\ref{fig:spatial-diffusion-x} for the position at $x=150\,\mathrm{pc}$. The spectrum is scaled with $p^{4.5}$, i.e. the initial power-law would be a horizontal line. At early times high-energy CRs can reach the position at $x=150\,\mathrm{pc}$ much faster, which results in larger amplitudes at high $p$. Over time, the amplitude of the low-energy CRs increases gradually, which reflects their longer diffusion times to the measurement position. At the same time the amplitude of the high-energy CRs starts to rapidly decrease because the high-energy component diffuses out of the domain and the total remaining high-energy component drops. Those two effects result in a temporal evolution of the spectrum that appears in a curved form with a maximum that shifts towards lower momenta. We note that locally in each bin the spectral slope remains the same as in the initial spectrum, $\slope=4.5$, because the diffusion coefficient scales with the momentum of the bin centre and we use the same diffusion coefficient for the number and the energy density. We would like to emphasise that the changes in the amplitudes of individual bins are solely due to the spatial diffusion. There is no transfer of number and energy density in momentum space, i.e. the spatial integral over each momentum bin is conserved separately. This also means that the diffusion of each momentum bin can be computed independently.

We additionally illustrate this effect in Fig.~\ref{fig:spatial-diffusion-3D} for two different times in a three-dimensional representation. The left-hand spectrum is measured after $10\,\mathrm{kyr}$, the right-hand counterpart after $100\,\mathrm{kyr}$. The red lines indicate the spectrum at $x=0$, the blue lines at $150\,\mathrm{pc}$. After $10\,\mathrm{kyr}$ the blue line clearly indicates that the spectrum is dominated by high-energy CRs -- we note the large dynamic range in $z$. At the origin the spectrum still appears to be flat. The decrease of the amplitude at $p\gtrsim1000\,\mathrm{GeV}/c$ is not visible (cf. black line in right-hand panel of Fig~\ref{fig:spatial-diffusion-x}). After $10\,\mathrm{kyr}$ the significantly faster diffusion for high $p$ is visible in the spectrum. By that time the diffusion of low-energy CRs results in a shallower distribution along $x$, which in turn yields a spectrum that develops a stronger curvature at higher momenta.

\section{Conclusions}
\label{sec:conclusions}

The spectral shape of CR protons is important to properly account for energy dependent losses, injection, spatial diffusion and CR ionisation and as a result an accurate modelling of the energetic impact of CR protons. Predictive comparisons to observations furthermore requires to follow the spectral shape of a CR population. As CR protons are dynamically relevant, we would like to model them with a full spectrum in every computational cell of a hydrodynamical simulation. This requires an efficient algorithm that adequately represents the CR proton spectrum spanning the relevant energy regime from low-energy CRs that cool via Coulomb losses up to high-energy CR that suffer hadronic losses with a minimum of spectral bins.

Here, we present a new method that solves the Fokker-Planck equation using a piecewise power-law representation of the particle distribution function. The implemented two-moment approach uses the number and energy density to compute the time evolution of the spectrum and does not rely on a continuous particle distribution function, which makes the code more versatile and stable. Because of the low number of spectral bins both the memory requirements and the computational cost are low in comparison to classical methods which require orders of magnitude more bins and adopt piecewise constant values. The method is therefore well suited to be coupled to hydrodynamics and solved together with the gas fluid dynamics in every cell of a three-dimensional hydrodynamical simulation.

For the adiabatic process the new scheme reveals orders of magnitude lower errors for the number and energy density and a very stable and accurate computation of the spectral slope. The combination of injection and cooling including Coulomb and hadronic losses shows very good agreement with theoretical steady state spectra. This method is also able to capture momentum-dependent CR diffusion, which causes a region outside the source to first see the highest-energy CRs before the low-energy CRs catch up.

Besides the dynamics a full spectral representation of CRs allows us to connect the CRs to observables like the chemical changes caused by low-energy CRs or the emission of $\gamma$-rays caused by hadronic interactions of high-energy CRs. In our follow-up paper, we explore the coupling of this new method to MHD to study the hydrodynamical impact of evolving the CR spectrum. 

\section*{Acknowledgements}
The authors thank Andy Strong and Stefanie Walch for fruitful discussions. We also thank the anonymous referee for very constructive comments that helped to improve the manuscript.
PG and TN acknowledge support from the DFG Priority Program 1573 \emph{Physics of the Interstellar Medium}.
PG and CP acknowledge funding from the European Research Council under ERC-CoG grant CRAGSMAN-646955.
MH acknowledges support of the (Polish) National Science Centre through the grant No. 2015/19/ST9/02959.
TN acknowledges support from the DFG cluster of excellence \emph{ORIGINS}.



\bibliographystyle{mnras}
\bibliography{astro,girichidis} 

\appendix

\section{Proof for the adiabatic process}
\label{sec:adiabatic-analytic-proof}

\begin{figure}
\includegraphics[width=8cm]{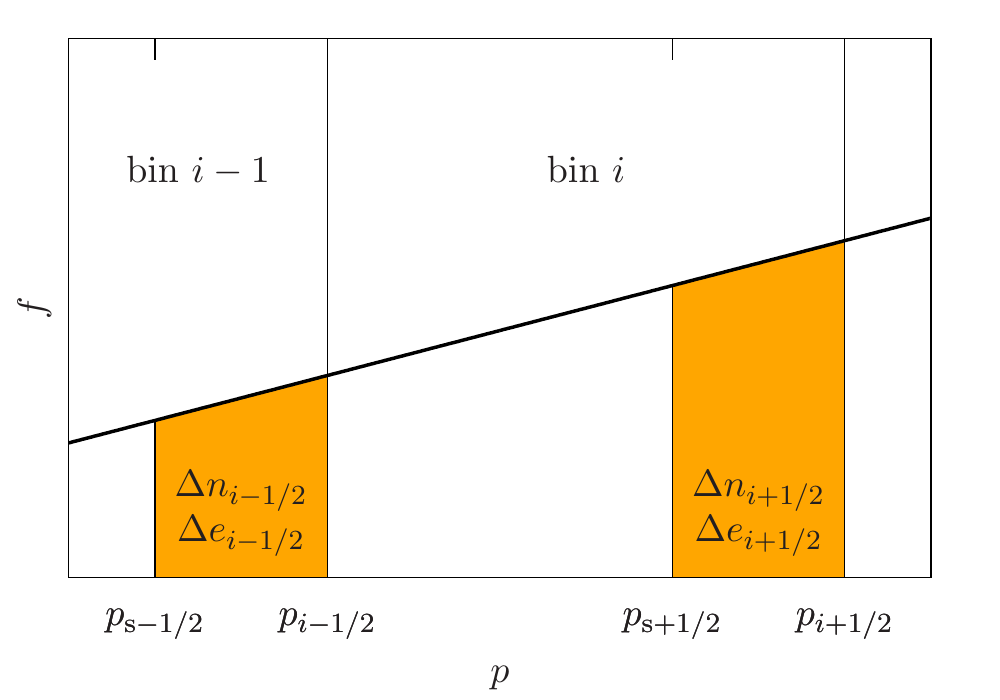}
\caption{Illustration of the momenta and transferred number and energy density for the analytic proof.}
\label{fig:adiabatic-compression-analytic-proof}
\end{figure}

The adiabatic process is simply an advection in (logarithmic) momentum space plus a change in normalisation. The slope of a power-law distribution, $\slope$, does not change under the impact of an adiabatic process. For our numerical scheme this means that the ratio $e_i/n_i$ does not change during compression or expansion. We thus take $f$ to be continuous with local slopes $\slope\equiv \slope_{i-1}=\slope_i=\slope_{i+1}$ as illustrated in Fig.~\ref{fig:adiabatic-compression-analytic-proof}. We note that this simple mathematical proof does not work for discontinuous $f$. This implies that $f_{i+1/2} = f_{i-1/2}(p_{i+1/2}/p_{i-1/2})^{-\slope}$. We make a few assumptions without loss of generality:
\begin{itemize}
\item We use the relativistic regime, so $T(p)=pc$, but the results are similarly true for all other power laws.
\item $\slope_i\neq4$ and $\slope\neq3$, so that we do not have to treat integrands scaling with $1/p$ separately.
\item We look at a compression ($p_{\mathrm{s}-1/2} < p_{i-1/2}$) with $p_{\mathrm{s}-1/2}$ and $p_{\mathrm{s}+1/2}$ for the left and right edge of bin $i$ and note that the ratios of the shift momenta at both boundaries of the bin $i$ are equal,
  \begin{equation}
    \frac{p_{\mathrm{s}-1/2}}{p_{i-1/2}}=\frac{p_{\mathrm{s}+1/2}}{p_{i+1/2}}.
  \end{equation}
\item By construction of a piecewise power-law distribution function and a continuous function we can write
\begin{equation}
f_{i+1/2} = f_{i-1/2}\left(\frac{p_{i+1/2}}{p_{i-1/2}}\right)^{-\slope}.
\end{equation}
\end{itemize}
We then need to show that
\begin{equation}
  \frac{e_i(t+\Delta t)}{n_i(t+\Delta t)} = \frac{e_i(t)}{n_i(t)}.
\end{equation}
The above simplifications give for the total number density in bin $i$
\begin{equation}
  n_i = 4\pi f_{i-1/2} \frac{p_{i-1/2}^3}{-\slope+3}\ekl{\rkl{\frac{p_{i+1/2}}{p_{i-1/2}}}^{-\slope+3}-1}.
\end{equation}
The integrated fluxes at $p_{i-1/2}$, $\Phi_n(p_{i-1/2})$, and at $p_{i+1/2}$, $\Phi_n(p_{i+1/2})$ lead to $\Delta n_{i-1/2}$ and $\Delta n_{i+1/2}$, which are given by
\begin{align}
  \Delta n_{i-1/2} &= 4\pi f_{i-1/2} \frac{p_{i-1/2}^3}{-\slope+3}\ekl{1-\rkl{\frac{p_{s-1/2}}{p_{i-1/2}}}^{-\slope+3}}\\
  \Delta n_{i+1/2} &= 4\pi f_{i+1/2} \frac{p_{i+1/2}^3}{-\slope+3}\ekl{1-\rkl{\frac{p_{s+1/2}}{p_{i+1/2}}}^{-\slope+3}}\\
  &= 4\pi\,f_{i-1/2} \rkl{\frac{p_{i+1/2}}{p_{i-1/2}}}^{-\slope}\, \frac{p_{i+1/2}^3}{-\slope+3}\times\notag\\
  &\qquad\ekl{1-\rkl{\frac{p_{s-1/2}}{p_{i-1/2}}}^{-\slope+3}}\notag\\
  &= 4\pi\,f_{i-1/2} \frac{p_{i-1/2}^3}{-\slope+3}\,\rkl{\frac{p_{i+1/2}}{p_{i-1/2}}}^{-\slope+3}\times\notag\\
  &\qquad\ekl{1-\rkl{\frac{p_{s-1/2}}{p_{i-1/2}}}^{-\slope+3}}\notag
\end{align}
and finally the new number density after compression can be reduced to 
\begin{align}
  n_i(t+\Delta t) &= n_i + \Delta n_{i-1/2} - \Delta n_{i+1/2}\\
  &= n_i \rkl{\frac{p_{s-1/2}}{p_{i-1/2}}}^{-\slope+3}
\end{align}
For the energy density we have the total energy in bin $i$
\begin{equation}
  e_i = 4\pi c f_{i-1/2} \frac{p_{i-1/2}^4}{-\slope+4}\ekl{\rkl{\frac{p_{i+1/2}}{p_{i-1/2}}}^{-\slope+4}-1},
\end{equation}
The total changes in the energy can be computed in two steps. The first is to compute the integrals over the distribution function from the shift momenta to the cell boundaries in the same way as for the number density. This is the part of the energy that is transferred between the bins. In a second step we need to consider that the total energy is shifted along the momentum axis and each particle with an energy density $e$ gains energy as
\begin{align}
T&\rightarrow T\,\frac{p_{i-1/2}}{p_{s-1/2}},
\end{align}
where we used the assumption that the particle has relativistic energies, i.e. $T=pc$. This applies to the entire energy bin. We therefore find the energy in bin $i$ as
\begin{align}
e_i(t+\Delta t) &= \frac{p_{i-1/2}}{p_{s-1/2}} \rkl{e_i + \Delta e_{i-1/2} - \Delta e_{i+1/2}}
\end{align}
For the first step we compute the integrals as
\begin{align}
  \Delta e_{i-1/2} &= \int_{p_{s-1/2}}^{p_{i-1/2}} 4\pi p^3 c f_{i-1/2}\rkl{\frac{p}{p_{i-1/2}}}^{-\slope} \dd p\\
  &= 4\pi c f_{i-1/2} \frac{p_{i-1/2}^4}{-\slope+4}\ekl{\rkl{\frac{p}{p_{i-1/2}}}^{-\slope+4}}_{p_{s-1/2}}^{p_{i-1/2}}\notag\\
  &= 4\pi c f_{i-1/2} \frac{p_{i-1/2}^4}{-\slope+4}\ekl{1-\rkl{\frac{p_{s-1/2}}{p_{i-1/2}}}^{-\slope+4}}\notag
\end{align}
and for the integral at the upper bin boundary
\begin{align}
  \Delta e_{i+1/2} &= 4\pi c f_{i+1/2} \frac{p_{i+1/2}^4}{-\slope+4}\ekl{1-\rkl{\frac{p_{s+1/2}}{p_{i+1/2}}}^{-\slope+4}}\\
  &= 4\pi c f_{i-1/2} \rkl{\frac{p_{i+1/2}}{p_{i-1/2}}}^{-\slope}\frac{p_{i+1/2}^4}{-\slope+4}\times\notag\\
  &\qquad\ekl{1-\rkl{\frac{p_{s-1/2}}{p_{i-1/2}}}^{-\slope+4}}\notag\\
  &= 4\pi c f_{i-1/2} \rkl{\frac{p_{i+1/2}}{p_{i-1/2}}}^{-\slope+4}\frac{p_{i-1/2}^4}{-\slope+4}\times\notag\\
  &\qquad\ekl{1-\rkl{\frac{p_{s-1/2}}{p_{i-1/2}}}^{-\slope+4}}\notag\\
    &= 4\pi c f_{i-1/2} \frac{p_{i-1/2}^4}{-\slope+4}\rkl{\frac{p_{i+1/2}}{p_{i-1/2}}}^{-\slope+4}\times\notag\\
   &\qquad\ekl{1-\rkl{\frac{p_{s-1/2}}{p_{i-1/2}}}^{-\slope+4}}\notag
\end{align}
where in the last equation we have used that $f_2 = f_1(p_2/p_1)^{-\slope}$ and $p_{s+1/2}/p_{i+1/2} = p_{s-1/2}/p_{i-1/2}$. The difference of those two terms gives
\begin{align}
  \Delta e_{i-1/2} - \Delta e_{i+1/2} &= e_i\ekl{\rkl{\frac{p_{s-1/2}}{p_{i-1/2}}}^{-\slope+4}-1}
\end{align}
Taken together, we find
\begin{align}
        e_{i} + \Delta e_{i-1/2} - \Delta e_{i+1/2} &=  e_i\,\rkl{\frac{p_{s-1/2}}{p_{i-1/2}}}^{-\slope+4}.
\end{align}
and finally
\begin{align}
e_i(t+\Delta t) &= e_i(t) \, \rkl{\frac{p_{s-1/2}}{p_{i-1/2}}}^{-\slope+3}
\end{align}
with the same scaling as the number density compared to initial value $e(t)$. The ratio after an adiabatic compression thus remains the same as well as the slope, $\slope$,
\begin{align}
\frac{e_i(t+\Delta t)}{n_i(t+\Delta t)} = \frac{e_i(t)}{n_i(t)},
\end{align}
which is required for an adiabatic change.

\bsp	
\label{lastpage}
\end{NoHyper}
\end{document}